\documentclass[twoside,preprintnumbers,amsmath,amssymb,pacs,nofootinbib]{revtex4}
\usepackage{graphicx}
\usepackage{dcolumn}
\usepackage{bm}
\usepackage{braket}
\usepackage{float}
\usepackage{txfonts}
\usepackage{pslatex}

\newcommand{\half}{\mbox{\small{$\frac{1}{2}$}}}
\newcommand{\MSbar}{\overline{\mbox{MS}}}

\newcommand{\p}{\partial}

\newcommand{\e}{\ensuremath{\mathrm{e}}}
\newcommand{\phys}{\ensuremath{\mathrm{phys}}}
\newcommand{\en}{\ensuremath{\mathrm{en}}}
\renewcommand{\d}{\ensuremath{\mathrm{d}}}
\newcommand{\Tr}{\ensuremath{\mathrm{Tr\;}}}
\newcommand{\lco}{\ensuremath{\overline{\varphi}\varphi-\overline{\omega}\omega}}
\newcommand{\lms}{\Lambda_{\overline{\mbox{\tiny{MS}}}}}
\newcommand{\omu}{\overline{\mu}}

\begin{document}
\title{{\Large A refinement of the Gribov-Zwanziger approach in the Landau gauge: infrared propagators in harmony with the lattice results}}

\author{D. Dudal$^a$}
    \email{david.dudal@ugent.be}

\author{J.~A.~Gracey$^b$}
    \email{gracey@liv.ac.uk}

\author{S.P. Sorella$^c$}
    \email{sorella@uerj.br}
    \altaffiliation{Work supported by FAPERJ, Funda{\c c}{\~a}o de Amparo {\`a} Pesquisa do Estado do Rio de Janeiro,
                                    under the program {\it Cientista do Nosso Estado}, E-26/100.615/2007.}
\author{N. Vandersickel$^a$}
    \email{nele.vandersickel@ugent.be}

\author{H. Verschelde$^a$}
 \email{henri.verschelde@ugent.be}
 \affiliation{\vskip 0.1cm
                            $^a$ Ghent University, Department of Mathematical Physics and Astronomy \\
                            Krijgslaan 281-S9, B-9000 Gent, Belgium\\\\\vskip 0.1cm
                            $^b$ Theoretical Physics Division, Department of Mathematical Sciences, University of Liverpool\\ P.O. Box 147, Liverpool, L69 3BX, United Kingdom \\\\\vskip 0.1cm
                            $^c$ Departamento de F\'{\i }sica Te\'{o}rica, Instituto de F\'{\i }sica, UERJ - Universidade do Estado do Rio de Janeiro\\
                            Rua S\~{a}o Francisco Xavier 524, 20550-013 Maracan\~{a}, Rio de Janeiro, Brasil
                            }


\begin{abstract}\noindent
Recent lattice data  have reported an infrared suppressed,
positivity violating gluon propagator which is nonvanishing
 at zero momentum and a ghost propagator which is no
longer enhanced. This paper discusses how to obtain analytical
results which are in qualitative agreement with these lattice data
within the Gribov-Zwanziger framework. This framework allows one
to take into account effects related to the existence of gauge
copies, by restricting the domain of integration  in
the path integral to the Gribov region. We elaborate to great
extent on a previous short paper by presenting additional results,
also confirmed by the numerical simulations.  A
detailed discussion on the soft breaking of the BRST symmetry
arising in the Gribov-Zwanziger approach is provided.

\end{abstract}

\preprint{LTH-789} \maketitle

\setcounter{page}{1}

\section{Introduction}
As is well known, quantum  chromodynamics (QCD) is confining at
low energy. Confinement means that it is impossible to detect free
quarks and gluons in the low momentum region as quarks form
colorless bound states like baryons and mesons.  Even if one omits
the quarks, pure $SU(N)$ Yang-Mills gauge theory remains confining
as gluons form bound states known as glueballs. Hitherto,
confinement is still poorly understood.  There is a widespread
belief that the infrared behavior of the gluon and ghost
propagator is deeply related to   the issue of
confinement and, therefore, these propagators have been widely
investigated. In this paper, we shall use the following
conventions for the gluon and the ghost propagator,
\begin{align}
\Braket{A_\mu^a(-p) A_\nu^b (p)} &=\delta^{ab}\mathcal{D}(p^2)\left(\delta_{\mu\nu}-\frac{p_\mu{p}_\nu}{p^2}\right)\;,  &\Braket{c^a(-p) \overline{c}^b(p)}&=\delta^{ab}\mathcal{G}(p^2)\;.
\end{align}
Until recently, lattice results have shown an infrared suppressed,
positivity violating gluon propagator which seemed to tend
towards zero for zero momentum, i.e. $\mathcal{D}(0) =0$, and a ghost propagator which was
believed to be enhanced in the infrared \cite{Cucchieri:2004mf,
Sternbeck:2004xr}, $\mathcal{G}(k^2 \approx 0) \sim 1/k^{2 + \kappa}$ with $\kappa > 0$. Different analytical approaches were in
agreement with these results (e.g.~\cite{Alkofer:2000wg,
Lerche:2002ep,
Pawlowski:2003hq,Alkofer:2003jj,Gribov:1977wm,Zwanziger:1989mf,Zwanziger:1992qr,
Zwanziger:2001kw}) to quote only a few).  For instance, several
works based on the Schwinger-Dyson or Exact Renormalization Group
equations reported an infrared enhanced ghost propagator and an
infrared suppressed, vanishing gluon propagator, obeying a power
law behavior characterized by a unique infrared exponent, as
stated by a sum rule discussed in \cite{Alkofer:2000wg,
Lerche:2002ep, Pawlowski:2003hq,Alkofer:2003jj}. The infrared propagators have also been studied from a thermodynamical viewpoint in \cite{Chernodub:2007rn}. Also the
Gribov-Zwanziger action predicts an infrared enhanced ghost
propagator and a zero-momentum vanishing gluon propagator
\cite{Zwanziger:1989mf,Zwanziger:1992qr}. This action was
constructed in order to analytically implement the restriction to
the Gribov region $\Omega$,   defined as the set of
field configurations fulfilling the Landau gauge condition and for
which the Faddeev-Popov operator,
\begin{eqnarray}
\mathcal{M}^{ab} &=&  -\partial_{\mu}\left( \p_{\mu} \delta^{ab} + g f^{acb} A^c_{\mu} \right) \;,
\end{eqnarray}
is strictly positive, namely
\begin{eqnarray}
\Omega &\equiv &\{ A^a_{\mu}, \; \partial_{\mu} A^a_{\mu}=0, \; \mathcal{M}^{ab}  >0\} \;.
\end{eqnarray}
The boundary, $\partial \Omega$, of the region $\Omega$ is called the (first) Gribov horizon. This restriction is necessary to avoid the appearance of Gribov copies in the Landau gauge related to
gauge transformations \cite{Gribov:1977wm}. However, this region $\Omega$ still contains a number of Gribov copies and is therefore still ``larger'' than the fundamental modular region (FMR), which is completely free of Gribov copies. Unfortunately, it is unknown how to treat the FMR analytically \cite{Semenov,Dell'Antonio:1991xt,Dell'Antonio2,vanBaal:1991zw}. \\

However, more recent lattice data
\cite{Cucchieri:2007rg,Cucchieri:2007md,Bogolubsky:2007ud,Cucchieri:2008fc}
at larger volumes display an infrared suppressed, positivity
violating gluon propagator, which is \textit{nonvanishing} at zero
momentum, , i.e. $\mathcal{D}(0) \not=0$, and a ghost propagator which is \textit{no longer
enhanced}, $\mathcal{G}(k^2 \approx 0) \sim 1/k^{2}$ . This implies that the previous mentioned analytical
approaches are not conclusive. It is worth pointing out that,
recently, the authors of \cite{Boucaud:2008ky,Aguilar:2008xm} have
obtained a solution of the Schwinger-Dyson equations which is in
agreement with the latest lattice data. Furthermore, as we have
shown in a previous work \cite{Dudal:2007cw}, this agreement can
also be found within the Gribov-Zwanziger approach. In this
framework, we have added a novel mass term to the original
Gribov-Zwanziger action. This new term corresponds to the
introduction of a dimension 2 operator. We recall that by
including condensates, which are the  vacuum expectation value of
certain local operators, one can take into account nonperturbative
effects  which play an important role in the infrared region.
During the course of the current work, it shall become clear that
we also have to add an additional vacuum term to
the action, which will allow us to stay within the Gribov
 region $\Omega$. The  previous paper
\cite{Dudal:2007cw} only gave a brief  account of the consequences
of adding the mass operator to the original Gribov-Zwanziger
action. For this reason, here we shall present an extensive study
of the Gribov-Zwanziger action with
the inclusion of the new parts.\\

The purpose of this paper is fourfold, and it is organized as
follows. The first aim, discussed in section II, is to give a
detailed proof of the renormalizability of the extended action.
Therefore, we first present an overview of the Gribov-Zwanziger
action, $S_{GZ}$,  in the Landau gauge which implements the
restriction the Gribov region $\Omega$. Next, we add the local
composite operator $S_m = \frac{m^2}{2} \int \d^4 x\; A_{\mu}^2$ to
this action and we prove the renormalizability of this extended
action, $S_{GZ} + S_m$. Subsequently, we show that by adding another
term, $S_M = M^2 \int
\d^4x\left[\left(\overline{\varphi}\varphi-\overline{\omega}\omega\right)
+ \frac{2 (N^2 -1)}{ g^2 N}   \varsigma  \lambda^2  \right]$, the
renormalizability is not destroyed. In summary, section II
establishes the renormalizability of the action $S_{GZ} + S_m +
S_M$. The second aim,  investigated in section III, is to
demonstrate that this extra term enables us to obtain propagators
which   exhibit the desired behavior. In particular, the tree level
gluon propagator is calculated explicitly and the ghost propagator
is determined up to one loop. Both the ghost and the gluon
propagator are in qualitative agreement with the latest lattice
results. Up to this point, we have added this mass term by hand.
Hence, a third aim is to obtain a dynamical value for $M^2$. Section
IV presents this dynamical value.   An estimate for the one loop
gluon propagator at zero momentum   as well as for the ghost
propagator at low momenta is given. Also, the positivity violation
of the gluon propagator is scrutinized and compared with the
available lattice data. The last aim is to highlight the BRST
breaking of the Gribov-Zwanziger action, which is presented in
detail in Section V. We already stress here that it is the
restriction to the Gribov region $\Omega$, implemented by the
Gribov-Zwanziger action, which induces the explicit breaking of the
BRST symmetry. Further, we provide a few remarks on the
Maggiore-Schaden approach to the issue of the BRST breaking
\cite{Maggiore:1993wq}, and we revisit a few aspects of the
Kugo-Ojima confinement criterion \cite{Kugo:1979gm}. We end this
paper with a discussion in section VI.

\section{The extended action and the renormalizability}
\subsection{The Gribov-Zwanziger action \label{GZ}}
We begin with an overview of the action constructed by Zwanziger
\cite{Zwanziger:2001kw} which implements the restriction to the
Gribov region $\Omega$  \cite{Gribov:1977wm} in
Euclidean Yang-Mills theories in the Landau gauge. We start from the
following action,
\begin{eqnarray}
S_h&=& S_{YM} + \int \d^{4}x\;\left( b^{a}\partial_\mu A_\mu^{a}
+\overline{c}^{a}\partial _{\mu } D_{\mu }^{ab}c^b \right) +
\gamma^4 \int \d^4 x\; h(x) \;,
\end{eqnarray}
with $S_{YM}$ the classical Yang-Mills action,
\begin{eqnarray}
S_{YM} &=& \frac{1}{4}\int \d^4 x F^a_{\mu\nu} F^a_{\mu\nu} \;,
\end{eqnarray}
and $h(x)$ the so called horizon function,
\begin{eqnarray}
 h(x) &=& g^2 f^{abc} A^b_{\mu} \left(\mathcal{M}^{-1}\right)^{ad} f^{dec} A^e_{\mu} \;.
\end{eqnarray}
The parameter $\gamma$, known as the Gribov parameter is not free and is determined by the horizon
condition:
\begin{eqnarray}\label{horizonconditon}
\braket{h(x)} &=& d (N^2 -1) \;,
\end{eqnarray}
 where $d$ is the number of space-time dimensions. The
nonlocal horizon function can be localized through a suitable set of
additional fields. The complete localized action reads
\begin{equation}
S=S_{0} +  S_{\gamma}\;, \label{s1eq1}
\end{equation}
with
\begin{eqnarray}\label{7}
S_{0} &=&S_{\mathrm{YM}}+\int \d^{4}x\;\left( b^{a}\partial_\mu A_\mu^{a}+\overline{c}^{a}\partial _{\mu } D_{\mu}^{ab}c^b \right) \;  \nonumber \\
&&+\int \d^{4}x\left( \overline{\varphi }_{\mu }^{ac}\partial _{\nu}\left(\partial _{\nu }\varphi _{\mu }^{ac}+gf^{abm}A_{\nu }^{b}\varphi _{\mu
}^{mc}\right) -\overline{\omega }_{\mu }^{ac}\partial _{\nu }\left( \partial_{\nu }\omega _{\mu }^{ac}+gf^{abm}A_{\nu }^{b}\omega _{\mu }^{mc}\right)
  -g\left( \partial _{\nu }\overline{\omega }_{\mu}^{ac}\right) f^{abm}\left( D_{\nu }c\right) ^{b}\varphi _{\mu
}^{mc}\right) \nonumber \;, \\
S_{\gamma}&=& -\gamma ^{2}g\int\d^{4}x\left( f^{abc}A_{\mu }^{a}\varphi _{\mu }^{bc}+f^{abc}A_{\mu}^{a}\overline{\varphi }_{\mu }^{bc} + \frac{4}{g}\left(N^{2}-1\right) \gamma^{2} \right) \;.
\end{eqnarray}
The fields $\left( \overline{\varphi }_{\mu
}^{ac},\varphi_{\mu}^{ac}\right) $ are a pair of complex conjugate
bosonic fields,   while $\left( \overline{\omega }_{\mu
}^{ac},\omega_{\mu}^{ac}\right) $ are anticommuting fields.  Each
of these fields has $4\left( N^{2}-1\right) ^{2}$ components. We
can easily see that the action $S_0$ displays a global $U(f)$
symmetry, $f=4\left(N^{2}-1\right) $, with respect to the
composite index $i=\left( \mu,c\right) =1,...,f$, of the
additional fields $\left( \overline{\varphi }_{\mu }^{ac},\varphi
_{\mu }^{ac},\overline{\omega }_{\mu}^{ac},\omega _{\mu
}^{ac}\right) $. Therefore, we simplify the notation of these
fields by setting
\begin{equation}
\left( \overline{\varphi }_{\mu }^{ac},\varphi _{\mu }^{ac},\overline{\omega}_{\mu }^{ac},\omega _{\mu }^{ac}\right) =\left( \overline{\varphi }_{i}^{a},\varphi _{i}^{a},\overline{\omega }_{i}^{a},\omega _{i}^{a}\right)\;,
\end{equation}
so we get
\begin{eqnarray}
S_{0} &=&S_{\mathrm{YM}}+\int \d^{4}x\;\left( b^{a}\partial_\mu A_\mu^{a}+\overline{c}^{a}\partial _{\mu }\left( D_{\mu }c\right) ^{a}\right) +\int \d^{4}x\left( \overline{\varphi }_{i}^{a}\partial _{\nu}\left( D_{\nu }\varphi _{i}\right) ^{a}-\overline{\omega}_{i}^{a}\partial _{\nu}\left( D_{\nu }\omega _{i}\right) ^{a} -g\left( \partial _{\nu }\overline{\omega }_{i}^{a}\right) f^{abm}\left( D_{\nu }c\right)^{b}\varphi _{i}^{m} \right) \;.\label{snul}
\end{eqnarray}

Now we shall try to translate the horizon condition
\eqref{horizonconditon} into a more practical version
\cite{Zwanziger:1992qr}. The local action  $S$ and the
nonlocal action $S_h$ are related as follows,
\begin{eqnarray}
    \int \d A \d b \d c \d \overline{c} \e^{-S_h} &=& \int \d A \d b \d c \d\overline{c} \d \varphi \d\overline{\varphi} \d \omega \d \overline{\omega} e^{-S} \;.
\end{eqnarray}
If we take the partial derivative of both sides with respect to $\gamma^2$ we obtain,
\begin{eqnarray}
-2 \gamma^2 \braket{h} &=& \braket{g f^{abc} A^a_{\mu} \varphi^{bc}_{\mu}} + \braket{g f^{abc} A^a_{\mu} \overline{\varphi}^{bc}_{\mu}} \;.
\end{eqnarray}
Using this last expression and assuming that $\gamma \not= 0$, we can rewrite the horizon condition \eqref{horizonconditon}
\begin{eqnarray}\label{horizonconditon2}
     \braket{g f^{abc} A^a_{\mu} \varphi^{bc}_{\mu}} + \braket{g f^{abc} A^a_{\mu} \overline{\varphi}^{bc}_{\mu}} + 2 \gamma^2 d (N^2 -1)   = 0 \;.
\end{eqnarray}
We know that the quantum action  $\Gamma$ is obtained through the definition
\begin{eqnarray}
\e^{-\Gamma} &=& \int \d\Phi \e^{-S} \;,
\end{eqnarray}
where $\int \d\Phi$ stands for the integration over all the fields.
It is now easy to see that
\begin{eqnarray}\label{gapgamma}
\frac{\p \Gamma}{\p \gamma^2} &=& 0
\end{eqnarray}
is exactly equivalent with equation \eqref{horizonconditon2}. Therefore, equation \eqref{gapgamma} represents the horizon condition. We
remark that the condition \eqref{gapgamma} also includes the solution $\gamma = 0$.  However, $\gamma=0$ would correspond to the case in which the
restriction to the Gribov region would not have been implemented. As such, the value $\gamma=0$ has to be disregarded as an artefact due to the reformulation of the horizon condition.  \\

As it has been proven in \cite{Zwanziger:1992qr}, the Gribov-Zwanziger action $S$ is renormalizable to all orders. In the next section, we shall give an overview of this renormalizability, but with the insertion of the local composite operator $A_{\mu }^{a}A_{\mu }^{a}$, to extend the action further. Obviously, the renormalizability of this extended action $S'$ also includes the renormalizability of the ordinary Gribov-Zwanziger action $S$.

\subsection{Adding the local composite operator $A_{\mu }^{a}A_{\mu}^{a}$}
If we add the local composite operator $A_{\mu}^{a}A_{\mu}^{a}$ to
 \eqref{s1eq1} one can prove \cite{Dudal:2005} that
 the following action is renormalizable to all orders
\begin{eqnarray}\label{actiemetA}
S' &=& S_0 + S_{\gamma} + S_{A^2},
\end{eqnarray}
with
\begin{eqnarray}
S_{A^2} &=& \int \d^{4}x\left( \frac{\tau }{2}A_{\mu }^{a}A_{\mu
}^{a}-\frac{\zeta }{2}\tau ^{2}\right) \label{s1eq3},
\end{eqnarray}
with $\tau$ a new source and $\zeta$ a new parameter. We now go a
little bit more into the details of the renormalization of this
action, as it will be useful later.  We remark that
if we prove the renormalizability of the action $S'$, we have also
proven the renormalizability of $S = S_0 + S_{\gamma}$ just by
putting $\tau$ equal to zero. We will use the method of algebraic
renormalization  \cite{Piguet:1995er}. Roughly speaking, this means that we will embed the action $S'$ into a
larger action by adding new sources, so it will display a greater
number of symmetries. These symmetries are important as they will
 imply constraints on the  possible allowed
counterterm. The larger the number of symmetries, the more
limitations we will find on the counterterm. This will lead to a
bigger possibility to absorb the counterterm into the original
action, thereby proving the renormalizability. In the end, we give
the sources the correct physical values, so we obtain the
action $S'$ again.\\

We shall now implement, step by step, this method of algebraic
renormalization. Firstly, we introduce two local external sources $M_{\mu}^{ai}$,\ $V_{\mu }^{ai}$
so we can treat $f^{abc}A_{\mu }^{a}\varphi_{\mu}^{bc}$ and
$f^{abc}A_{\mu }^{a}\overline{\varphi }_{\mu}^{bc}$ as composite
operators just like $A^2_{\mu}$. Hence, we replace the term $S_{\gamma}$ by
\begin{eqnarray}
S'_{\gamma} &=& -\int \d^{4}x\left( M_{\mu }^{ai}\left( D_{\mu}\varphi _{i}\right) ^{a}+V_{\mu }^{ai}\left( D_{\mu}\overline{\varphi }_{i}\right) ^{a} + 4 \gamma^4 (N^2 -1)\right) \;. \label{s1eq5}
\end{eqnarray}
If we set the sources to their physical values in the
end
\begin{equation}\label{physval2}
\left. M_{\mu \nu }^{ab}\right|_{\phys}= \left.V_{\mu \nu}^{ab}\right|_{\phys}=\gamma ^{2}\delta ^{ab}\delta _{\mu \nu }\;,
\end{equation}
we obtain, as requested, the term $S_{\gamma}$ defined in \eqref{s1eq1}.\\
Secondly, the algebraic renormalization procedure
requires this action to be BRST invariant. Therefore, we further
introduce three extra sources $N_{\mu}^{ai}$,\ $U_{\mu }^{ai}$ and
$\eta$ and replace $S'_{\gamma} + S_{A^2}$ by
\begin{eqnarray}
S_{\mathrm{s}} &=& s\int \d^{4}x\left( -U_{\mu }^{ai}\left( D_{\mu
}\varphi _{i}\right) ^{a}-V_{\mu }^{ai}\left( D_{\mu
}\overline{\omega }_{i}\right) ^{a}-U_{\mu
}^{ai}V_{\mu }^{ai}+\frac{1}{2}\eta A_{\mu }^{a}A_{\mu }^{a}-\frac{1}{2}%
\zeta \tau \eta \right)\nonumber\\
&=& \int \d^{4}x\left( -M_{\mu }^{ai}\left( D_{\mu }\varphi
_{i}\right) ^{a}-gU_{\mu }^{ai}f^{abc}\left( D_{\mu }c\right)
^{b}\varphi _{i}^{c}+U_{\mu }^{ai}\left( D_{\mu }\omega _{i}\right)
^{b}\right.
\nonumber \\
&& - N_{\mu }^{ai}\left( D_{\mu }\overline{\omega }_{i}\right)
^{a}-V_{\mu }^{ai}\left( D_{\mu }\overline{\varphi }_{i}\right)
^{a}+gV_{\mu }^{ai}f^{abc}\left( D_{\mu }c\right)
^{b}\overline{\omega }_{i}^{c}
\nonumber \\
 &-&\left.M_{\mu }^{ai}V_{\mu }^{ai}+U_{\mu }^{ai}N_{\mu }^{ai}+\frac{1}{2}%
\tau A_{\mu }^{a}A_{\mu }^{a}+\eta A_{\mu }^{a}\partial _{\mu }c^{a}-\frac{1%
}{2}\zeta \tau ^{2}\right) \;,
\end{eqnarray}
where the BRST  transformations of all the fields and
sources are:
\begin{align}\label{BRST}
sA_{\mu }^{a} &=-\left( D_{\mu }c\right) ^{a}\;, & sc^{a} &=\frac{1}{2}gf^{abc}c^{b}c^{c}\;,   \nonumber \\
s\overline{c}^{a} &=b^{a}\;,&   sb^{a}&=0\;,  \nonumber \\
s\varphi _{i}^{a} &=\omega _{i}^{a}\;,&s\omega _{i}^{a}&=0\;,\nonumber \\
s\overline{\omega }_{i}^{a} &=\overline{\varphi }_{i}^{a}\;,& s
\overline{\varphi }_{i}^{a}&=0\;,
\end{align}
and
\begin{align}\label{BRSTbis}
sU_{\mu }^{ai} &= M_{\mu }^{ai}\;, & sM_{\mu }^{ai}&=0\;,  \nonumber \\
sV_{\mu }^{ai} &= N_{\mu }^{ai}\;, & sN_{\mu }^{ai}&=0\;,  \nonumber\\
s\eta &=\tau \;,&s\tau &=0\;.
\end{align}
We recall that the BRST  operator $s$ is nilpotent,
meaning that $s^2 = 0$. We mention again that by replacing the
sources with their physical values in the end
\begin{eqnarray}
&\left. U_{\mu }^{ai}\right|_{\phys} = \left. N_{\mu}^{ai}\right|_{\phys} = 0 \;,& \label{physval1}\\
 &\left. \eta \right|_{\phys} = 0\;,&
\end{eqnarray}
 one recovers the original terms $S_{\gamma} + S_{A^2}$.\\
Finally, a term $S_{\mathrm{ext}}$,
\begin{eqnarray}
S_{\mathrm{ext}}&=&\int \d^{4}x\left( -K_{\mu }^{a}\left( D_{\mu }c\right) ^{a}+\frac{1}{2}gL^{a}f^{abc}c^{b}c^{c}\right) \;,
\end{eqnarray}
was added, which is needed to define the nonlinear BRST transformations of the
gauge and ghost fields. $K_{\mu }^{a}$ and $L^{a}$ are two new sources,
invariant under the BRST symmetry $s$ and with
\begin{eqnarray}
\left. K_{\mu }^{a}\right|_{\phys} =\left. L^{a}\right|_{\phys}  = 0\;.&
\end{eqnarray}
The enlarged action is thus given by
\begin{equation}\label{enlarged}
\Sigma =S_{0}+S_{\mathrm{s}}+S_{\mathrm{ext}}\;,
\end{equation}
and one easily sees that the action $\Sigma$ is indeed BRST
invariant. This action now enjoys  a larger number of Ward
identities summarized  as follows:
\begin{table}[t]
  \centering
        \begin{tabular}{|c|c|c|c|c|c|c|c|c|}
        \hline
        & $A_{\mu }^{a}$ & $c^{a}$ & $\overline{c}^{a}$ & $b^{a}$ & $\varphi_{i}^{a} $ & $\overline{\varphi }_{i}^{a}$ &                $\omega _{i}^{a}$ & $\overline{\omega }_{i}^{a}$ \\
        \hline
        \hline
        \textrm{dimension} & $1$ & $0$ &$2$ & $2$ & $1$ & $1$ & $1$ & $1$ \\
        \hline
        $\mathrm{ghost\; number}$ & $0$ & $1$ & $-1$ & $0$ & $0$ & $0$ & $1$ & $-1$ \\
        \hline
        $Q_{f}\textrm{-charge}$ & $0$ & $0$ & $0$ & $0$ & $1$ & $-1$& $1$ & $-1$\\
        \hline
        \end{tabular}
        \caption{Quantum numbers of the fields.}\label{tabel1}
        \end{table}
        \begin{table}[t]
    \centering
    \begin{tabular}{|c|c|c|c|c|c|c|c|c|}
        \hline
        &$U_{\mu}^{ai}$&$M_{\mu }^{ai}$&$N_{\mu }^{ai}$&$V_{\mu }^{ai}$&$K_{\mu }^{a}$&$L^{a}$& $\tau$& $\eta$  \\
        \hline
        \hline
        \textrm{dimension} & $2$ & $2$ & $2$ &$2$  & $3$ & $4$ & $2$ & $2$  \\
        \hline
        $\mathrm{ghost\; number}$ & $-1$& $0$ & $1$ & $0$ & $-1$ & $-2$ & 0 & 0 \\
        \hline
        $Q_{f}\textrm{-charge}$ & $-1$ & $-1$ & $1$ & $1$ & $0$ & $0$  & 0 & 0 \\
        \hline
        \end{tabular}
        \caption{Quantum numbers of the sources.}\label{tabel2}
        \end{table}
\begin{itemize}
\item For the $U(f)$ invariance mentioned before we have
\begin{eqnarray}
U_{ij} \Sigma &=&0\;,\nonumber\\
  U_{ij}&=&\int \d^{4}x\left( \varphi
_{i}^{a}\frac{\delta }{\delta \varphi _{j}^{a}}-\overline{\varphi
}_{j}^{a}\frac{\delta }{\delta \overline{\varphi
}_{i}^{a}}+\omega _{i}^{a}\frac{\delta }{\delta \omega _{j}^{a}}-\overline{\omega }_{j}^{a}\frac{\delta }{\delta \overline{\omega }_{i}^{a}}
+  M^{ai}_{\mu} \frac{\delta}{\delta M^{aj}_{\mu}} -
U^{aj}_{\mu}\frac{\delta}{\delta U^{ai}_{\mu}} +
N^{ai}_{\mu}\frac{\delta}{\delta N^{aj}_{\mu}} -
V^{aj}_{\mu}\frac{\delta}{\delta V^{ai}_{\mu}} \right) \;.
\label{ward1}
\end{eqnarray}
By means of the diagonal operator $Q_{f}=U_{ii}$, the
$i$-valued fields and sources turn out to possess an additional quantum number.
One can find all quantum numbers in TABLE \ref{tabel1} and TABLE \ref{tabel2}.
\item  The Slavnov-Taylor identity reads
\begin{equation}
\mathcal{S}(\Sigma )=0\;,
\end{equation}
with
\begin{eqnarray}\label{slavnov}
\mathcal{S}(\Sigma ) &=&\int \d^{4}x\left( \frac{\delta \Sigma
}{\delta K_{\mu }^{a}}\frac{\delta \Sigma }{\delta A_{\mu
}^{a}}+\frac{\delta \Sigma }{\delta L^{a}}\frac{\delta \Sigma
}{\delta c^{a}}+b^{a}\frac{\delta \Sigma
}{\delta \overline{c}^{a}}+\overline{\varphi }_{i}^{a}\frac{\delta \Sigma }{\delta \overline{\omega }_{i}^{a}}+\omega _{i}^{a}\frac{\delta \Sigma }{\delta \varphi _{i}^{a}}+M_{\mu }^{ai}\frac{\delta \Sigma
}{\delta U_{\mu}^{ai}}+N_{\mu }^{ai}\frac{\delta \Sigma }{\delta V_{\mu }^{ai}}\right) \;.
\end{eqnarray}

\item  The Landau gauge condition and the antighost equation are given by
\begin{eqnarray}
\label{gaugeward}\frac{\delta \Sigma }{\delta b^{a}}&=&\partial_\mu A_\mu^{a}\;,\\
\frac{\delta \Sigma }{\delta \overline{c}^{a}}+\partial _{\mu
}\frac{\delta \Sigma }{\delta K_{\mu }^{a}}&=&0\;.
\end{eqnarray}

\item  The ghost Ward identity is
\begin{equation}
\mathcal{G}^{a}\Sigma =\Delta _{\mathrm{cl}}^{a}\;,
\end{equation}
with
\begin{eqnarray}
\mathcal{G}^{a} &=&\int \d^{4}x\left( \frac{\delta }{\delta c^{a}}+gf^{abc}\left( \overline{c}^{b}\frac{\delta }{\delta b^{c}}+\varphi _{i}^{b}\frac{\delta }{\delta \omega _{i}^{c}}+\overline{\omega }_{i}^{b}\frac{\delta }{\delta \overline{\varphi }_{i}^{c}}+V_{\mu }^{bi}\frac{\delta }{\delta N_{\mu }^{ci}}+U_{\mu }^{bi}\frac{\delta }{\delta M_{\mu }^{ci}}\right) \right) \;,  \nonumber \\
&&
\end{eqnarray}
and
\begin{equation}
\Delta _{\mathrm{cl}}^{a}=g\int \d^{4}xf^{abc}\left( K_{\mu}^{b}A_{\mu }^{c}-L^{b}c^{c}\right) \;.
\end{equation}
Notice that the term $\Delta _{\mathrm{cl}}^{a}$, being linear in
the quantum fields $A_{\mu }^{a}$, $c^{a}$, is a classical breaking.

\item  The linearly broken local constraints yield
\begin{equation}
\frac{\delta \Sigma }{\delta \overline{\varphi }^{ai}}+\partial _{\mu }\frac{\delta \Sigma }{\delta M_{\mu }^{ai}}=gf^{abc}A_{\mu }^{b}V_{\mu}^{ci} \;,
\end{equation}
\begin{equation}
\frac{\delta \Sigma }{\delta \omega ^{ai}}+\partial _{\mu}\frac{\delta \Sigma }{\delta N_{\mu
}^{ai}}-gf^{abc}\overline{\omega }^{bi}\frac{\delta \Sigma }{\delta b^{c}}=gf^{abc}A_{\mu }^{b}U_{\mu }^{ci} \;,
\end{equation}
\begin{equation}
\frac{\delta \Sigma }{\delta \overline{\omega }^{ai}}+\partial _{\mu }\frac{\delta \Sigma }{\delta U_{\mu }^{ai}}-gf^{abc}V_{\mu
}^{bi}\frac{\delta \Sigma }{\delta K_{\mu }^{c}}=-gf^{abc}A_{\mu
}^{b}N_{\mu }^{ci}  \;,
\end{equation}
\begin{equation}
\frac{\delta \Sigma }{\delta \varphi ^{ai}}+\partial _{\mu}\frac{\delta \Sigma }{\delta V_{\mu
}^{ai}}-gf^{abc}\overline{\varphi }^{bi}\frac{\delta \Sigma }{\delta b^{c}}-gf^{abc}\overline{\omega }^{bi}\frac{\delta \Sigma }{\delta \overline{c}^{c}}-gf^{abc}U_{\mu }^{bi}\frac{\delta \Sigma}{\delta K_{\mu }^{c}} =gf^{abc}A_{\mu }^{b}M_{\mu }^{ci}  \;.
\end{equation}

\item  The exact $\mathcal{R}_{ij}$ symmetry reads
\begin{equation}
\mathcal{R}_{ij}\Sigma =0\;,
\end{equation}
with
\begin{equation}
\mathcal{R}_{ij}=\int \d^{4}x\left( \varphi _{i}^{a}\frac{\delta}{\delta\omega _{j}^{a}}-\overline{\omega }_{j}^{a}\frac{\delta }{\delta \overline{\varphi }_{i}^{a}}+V_{\mu }^{ai}\frac{\delta }{\delta N_{\mu
}^{ai}}-U_{\mu }^{ai}\frac{\delta }{\delta M_{\mu }^{ai}}\right) \;.\label{ward7}
\end{equation}
\end{itemize}

When we turn to the quantum  level, we can use these
symmetries to characterize the most general  allowed
invariant counterterm $\Sigma ^{c}$. Following the algebraic
renormalization procedure \cite{Piguet:1995er}, $\Sigma ^{c}$ is
an integrated local polynomial in the fields and sources with
dimension bounded by four, and with vanishing ghost number and
$Q_{f}$-charge. The  previous Ward identities imply the
following constraints for $\Sigma^c$:
\begin{itemize}
\item The $U(f)$ invariance:
\begin{eqnarray}\label{conditiebegin}
U_{ij} \Sigma^c &=&0 \;.
\end{eqnarray}

\item The linearized Slavnov-Taylor identity:
\begin{equation}
\mathcal{B}_{\Sigma }\Sigma ^{c}=0\;,
\end{equation}
with $\mathcal{B}_{\Sigma }$ the nilpotent linearized Slavnov-Taylor operator,
\begin{eqnarray}
\mathcal{B}_{\Sigma } &=&\int \d^{4}x\left( \frac{\delta \Sigma}{\delta K_{\mu }^{a}}\frac{\delta }{\delta A_{\mu }^{a}}+\frac{\delta \Sigma }{\delta A_{\mu }^{a}}\frac{\delta }{\delta K_{\mu }^{a}}+\frac{\delta
\Sigma }{\delta L^{a}}\frac{\delta }{\delta c^{a}}+\frac{\delta\Sigma }{\delta c^{a}}\frac{\delta }{\delta L^{a}}+b^{a}\frac{\delta }{\delta \overline{c}^{a}}+\overline{\varphi}_{i}^{a}\frac{\delta }{\delta \overline{\omega }_{i}^{a}}+\omega_{i}^{a}\frac{\delta }{\delta \varphi_{i}^{a}}+M_{\mu }^{ai}\frac{\delta }{\delta U_{\mu }^{ai}}+N_{\mu }^{ai}\frac{\delta }{\delta V_{\mu }^{ai}} \right)\,,
\end{eqnarray}
and
\begin{equation}
\mathcal{B}_{\Sigma }\mathcal{B}_{\Sigma }=0\;.
\end{equation}

\item The Landau gauge condition and the antighost equation:
\begin{eqnarray}
\frac{\delta \Sigma ^{c}}{\delta b^{a}} &=&0\;, \nonumber \\
\frac{\delta \Sigma }{\delta \overline{c}^{a}}+\partial
_{\mu}\frac{\delta\Sigma }{\delta K_{\mu }^{a}} &=&0\;.
\end{eqnarray}

\item The ghost Ward identity:
\begin{equation}
\mathcal{G}^{a}\Sigma ^{c}=0\,.
\end{equation}

\item  The linearly broken local constraints:
\begin{eqnarray}
\frac{\delta \Sigma ^{c}}{\delta \varphi ^{ai}}+\partial _{\mu}\frac{\delta\Sigma ^{c}}{\delta V_{\mu }^{ai}}-gf^{abc}\overline{\omega }^{bi}\frac{\delta \Sigma ^{c}}{\delta \overline{c}^{c}}-gf^{abc}U_{\mu }^{bi}\frac{\delta \Sigma ^{c}}{\delta K_{\mu }^{c}} &=&0\;,  \nonumber \\
\frac{\delta \Sigma ^{c}}{\delta \overline{\omega }^{ai}}+\partial _{\mu }\frac{\delta \Sigma ^{c}}{\delta U_{\mu }^{ai}}-gf^{abc}V_{\mu }^{bi}\frac{\delta \Sigma ^{c}}{\delta K_{\mu }^{c}} &=&0\;,  \nonumber \\
\frac{\delta \Sigma ^{c}}{\delta \omega ^{ai}}+\partial _{\mu}\frac{\delta \Sigma ^{c}}{\delta N_{\mu }^{ai}} &=&0\;,\nonumber\\
\frac{\delta \Sigma }{\delta \overline{\varphi }^{ai}}+\partial _{\mu }\frac{\delta \Sigma }{\delta M_{\mu }^{ai}} &=&0\;.  \nonumber
\end{eqnarray}

\item The exact $\mathcal{R}_{ij}$ symmetry:
\begin{equation}\label{conditieeind}
\mathcal{R}_{ij}\Sigma ^{c}=0\;.
\end{equation}
\end{itemize}

\noindent These constraints imply that $\Sigma^{c}$ does not depend on the Lagrange multiplier $b^{a}$, and that the antighost $\overline{c}^{a}$ and the $i$-valued fields $\varphi_{i}^{a}$, $\omega _{i}^{a}$, $\overline{\varphi }_{i}^{a}$, $\overline{\omega }_{i}^{a}$ can enter only through the combinations \cite{Zwanziger:1992qr, Dudal:2005}
\begin{eqnarray}\label{combinatiesmogelijk}
\widetilde{K}_{\mu }^{a} &=&K_{\mu }^{a}+\partial _{\mu }\overline{c}^{a}-gf^{abc}\widetilde{U}_{\mu }^{bi}\varphi ^{ci}-gf^{abc}V_{\mu }^{bi}\overline{\omega }^{ci}\;,  \nonumber \\
\widetilde{U}_{\mu }^{ai} &=&U_{\mu }^{ai}+\partial _{\mu }\overline{\omega }^{ai}\;,  \nonumber \\
\widetilde{V}_{\mu }^{ai} &=&V_{\mu }^{ai}+\partial _{\mu }\varphi^{ai}\;,\nonumber \\
\widetilde{N}_{\mu }^{ai} &=&N_{\mu }^{ai}+\partial _{\mu }\omega^{ai}\;,
\nonumber \\
\widetilde{M}_{\mu }^{ai} &=&M_{\mu }^{ai}+\partial
_{\mu}\overline{\varphi }^{ai}\;.
\end{eqnarray}
The most general counterterm fulfilling the conditions \eqref{conditiebegin} - \eqref{conditieeind} contains four arbitrary parameters, $a_{0}$, $a_{1}$, $a_{2}$, $a_{3}$ and reads
\begin{eqnarray}
\Sigma ^{c} =a_{0}S_{YM}&+&a_{1}\int \d^{4}x\left( A_{\mu
}^{a}\frac{\delta
S_{YM}}{\delta A_{\mu }^{a}}+\widetilde{K}_{\mu }^{a}\partial _{\mu }c^{a}+
\widetilde{V}_{\mu }^{ai}\widetilde{M}_{\mu }^{ai}-\widetilde{U}_{\mu }^{ai}
\widetilde{N}_{\mu }^{ai}\right)+\int \d^{4}x\left( \frac{a_{2}}{2}\tau A_{\mu }^{a}A_{\mu }^{a}+
\frac{a_{3}}{2}\zeta \tau ^{2}+\left( a_{2}-a_{1}\right) \eta A_{\mu
}^{a}\partial _{\mu }c^{a}\right) \;.
\end{eqnarray}
Once the most general counterterm  has been determined,
one can straightforwardly verify that it can be reabsorbed through
 a multiplicative renormalization of the fields,
sources and coupling constants. We also mention the renormalization
 factors, useful for later calculations. If we set $\phi
=(A_{\mu }^{a}$, $c^{a}$, $\overline{c}^{a}$, $b^{a}$, $\varphi
_{i}^{a}$, $\omega _{i}^{a}$, $\overline{\varphi }_{i}^{a}$,
$\overline{\omega }_{i}^{a}) $ for all the fields and $\Phi
=(K^{a\mu }$, $L^{a}$, $M_{\mu }^{ai}$, $N_{\mu }^{ai}$, $V_{\mu
}^{ai}$, $U_{\mu }^{ai},\tau $, $\eta )$ for the sources, and if we
define
\begin{align}
g_{0}&=Z_{g}g\;, & \zeta _{0}&=Z_{\zeta }\zeta \;, \nonumber\\
 \phi _{0} &=Z_{\phi}^{1/2}\phi \;, & \Phi _{0} &=Z_{\Phi }\Phi \;,
\end{align}
one can determine
\begin{eqnarray}
Z_{g} &=& 1+\eta \frac{a_{0}}{2} \;,  \nonumber \\
Z_{A}^{1/2} &=& 1+\eta \left( a_{1}-\frac{a_{0}}{2}\right)  \;, \nonumber\\
Z_{\zeta }&=&1+\eta (-a_{3}-2a_{2}+4a_{1}-2a_{0}) \;.
\end{eqnarray}
These are the only independent renormalization constants. For example, the Faddeev-Popov ghosts $\left( c^{a},\overline{c}^{a}\right) $ and the $i$-valued fields $\left( \varphi _{i}^{a},\omega _{i}^{a},\overline{\varphi }_{i}^{a}, \overline{\omega }_{i}^{a}\right) $ have a common renormalization constant, determined by the renormalization constants $Z_{g}$ and $Z_{A}^{1/2}$,
\begin{equation}\label{wardid}
Z_{c}=Z_{\overline{c}}=Z_{\varphi }=Z_{\overline{\varphi }}=Z_{\omega }=Z_{\overline{\omega }}=\left( 1-\eta a_{0}\right)
=Z_{g}^{-1}Z_{A}^{-1/2}\;.
\end{equation}
The renormalization of the sources $\left( M_{\mu}^{ai},N_{\mu }^{ai},V_{\mu }^{ai},U_{\mu }^{ai}\right) $ is also
determined by the renormalization constants $Z_{g}$ and $Z_{A}^{1/2}$, being given by
\begin{equation}\label{zm}
Z_{\gamma^2}\equiv
Z_{M}=Z_{N}=Z_{V}=Z_{U}=Z_{g}^{-1/2}Z_{A}^{-1/4}\;.
\end{equation}
Also $Z_{\tau}$ is related to $Z_{g}$ and $Z_{A}^{1/2}$ \cite{Dudal:2005}:
\begin{eqnarray}
    Z_\tau=Z_gZ_A^{-1/2}\;.
\end{eqnarray}
Finally, $Z_b$, $Z_K$ and $Z_L$ are also not independent as they are given by:
\begin{align}
Z_b &= Z_A^{-1}\;, &
Z_K &= Z_c^{1/2}\;, &
Z_L&= Z_{A}^{1/2}\;.
\end{align}

\subsection{Adding a new mass term}
\subsubsection{Extended action}
We first explain the need for the inclusion of a new
dynamical effect. According to the latest lattice results, the gluon
propagator does not seem to vanish for zero momentum. This is
incompatible with the actions \eqref{s1eq1} and \eqref{actiemetA},
which both lead to a vanishing gluon propagator near the origin.
The tree level gluon propagator in the Gribov-Zwanziger model reads
\cite{Dudal:2005}
\begin{equation}
\Braket{ A_\mu^a(-p)A_\nu^b(p)} \equiv
\delta^{ab}\mathcal{D}(p^2)\left(\delta_{\mu\nu}-
\frac{p_\mu{p}_\nu}{p^2}\right)=\delta^{ab}\frac{p^2}{p^4+\lambda^4}\left(\delta_{\mu\nu}-
\frac{p_\mu{p}_\nu}{p^2}\right)\;,\label{propagatorgrbov}
\end{equation}
where we have set
\begin{equation}
\lambda^4 = 2g^2 N \gamma^4 \;. \label{lambda4}
\end{equation}
One recognizes indeed that expression \eqref{propagatorgrbov}
vanishes at the origin due to the presence of Gribov parameter
$\lambda$. In the $A^2_{\mu}$ model the gluon propagator
is modified in the following form,
\begin{equation}
 \Braket{A_\mu^a(-p)A_\nu^b(p)} \equiv \delta^{ab} \mathcal{D}(p^2) \left(\delta_{\mu\nu}-\frac{p_\mu{p}_\nu}{p^2}\right)=\delta^{ab}\frac{p^2}{p^4+m^2p^2+\lambda^4}\left(\delta_{\mu\nu}-
\frac{p_\mu{p}_\nu}{p^2}\right)\;, \label{propagatorAkwadraat}
\end{equation}
which reveals a further suppression  near the origin and thus it
still vanishes. We recall here that the fields $\left( \overline{
\varphi }_{\mu }^{ac},\varphi _{\mu }^{ac},\overline{\omega
}_{\mu}^{ac},\omega _{\mu }^{ac}\right)$ were introduced to localize
the horizon function \cite{Zwanziger:1992qr}, which implements the
restriction to the Gribov-region  $\Omega$. If we take a closer look
at the action \eqref{snul}, we observe an $A\varphi$-coupling at the
quadratic level. One can suspect that a nontrivial effect in the
$\varphi$-sector will immediately get translated into the gluon
sector.  For this reason, if we try to give a mass to the
$\overline{\varphi} \;,\varphi$-fields without spoiling the
renormalizability of the action, we might be able to modify the
gluon propagator in the desired way. Implementing this idea means
that we add a new term to the action \eqref{actiemetA} of the form
$J\overline{\varphi}^{a}_i \varphi^{a}_i$, with $J$ a new source. If
we want to preserve the renormalizability we have to add the mass
term in a BRST invariant way. Therefore, we consider the following
extended action:
\begin{eqnarray}\label{extended_action}
    S'' &=& S' + S_{\overline{\varphi} \varphi}\;, \\
    S_{\overline{\varphi} \varphi} &=& \int \d^4 x \left( s(-J \overline{\omega}^a_i \varphi^a_{i}) + \rho J \tau \right)    \nonumber\\
    &=&\int \d^4 x\left( -J\left( \overline{\varphi}^a_i \varphi^a_{i} - \overline{\omega}^a_i \omega^a_i \right) + \rho J \tau   \right) \;,
\end{eqnarray}
with $\rho$ a parameter and $J$ a dimension two source, invariant under the BRST transformation
\begin{align}
    sJ &= 0\;.
\end{align}

\subsubsection{Renormalizability}
The proof of the renormalizability of this action $S^{\prime
\prime}$ can  be easily done with the help of  the
Ward identities derived in the previous section. Again, we embed
the action $S''$ into a larger action,
\begin{eqnarray}\label{61}
\Sigma' &=& \Sigma + S_{\overline{\varphi} \varphi} \;,
\end{eqnarray}
containing more symmetries. It is subsequently trivial to check that all Ward identities \eqref{ward1}-\eqref{ward7} remain
unchanged up to potential harmless linear breaking terms and therefore the constraints \eqref{conditiebegin}-\eqref{conditieeind} as well as the combinations \eqref{combinatiesmogelijk} are preserved. This implies
that the counterterm  $\Sigma^{c\prime}$
corresponding to the action $\Sigma'$ is now given by
\begin{eqnarray} \label{tegenterm}
\Sigma^{c\prime} &=& \Sigma^c + \Sigma^c_{\varphi\overline{\varphi}} \;, \nonumber\\
 \Sigma^c_{\varphi\overline{\varphi}} &=&  a_4 J \tau \;,
\end{eqnarray}
with $a_4$ an arbitrary parameter. This counterterm can be absorbed
into the original action $\Sigma'$, hence we have proven the
renormalizability of our extended action. If we define
\begin{align}
   J_{0} &= Z_{J} J \;,  &  \rho_{0} &= Z_{\rho} \rho  \;,
\end{align}
we find
\begin{align}
Z_J &= Z_{\varphi}^{-1} = Z_g Z_A^{1/2}\;, &    Z_{\rho} &= 1 + \eta (a_4 - \frac{a_0}{2} - a_2)\;.
\end{align}

As the reader might have noticed, symmetries do also not prevent a
term $\kappa J^2$ to occur, with $\kappa$ a new parameter, but
we can  argue that $\kappa$ is in fact a redundant parameter, as no divergences in $J^2$ will occur. A term of this
form is independent of the fields, hence  it would only
be necessary to get rid of the infinities in the functional
energy, which we calculate by integrating the action over all the
fields
\begin{equation}
\int \d \Phi \e^{-S^{\prime\prime}} = \e^{-W(J)}\;.
\end{equation}
Seen from another perspective, we need a counterterm $\propto J^2$
to remove possible divergences in the vacuum correlators
$\Braket{\bigl(\overline{\varphi}\varphi-\overline{\omega}\omega\bigr)_x\bigl(\overline{\varphi}\varphi-\overline{\omega}\omega\bigr)_y}$
for $x\to y$. Such new divergences are typical when a local
composite operator (LCO) of dimension 2 is added to the theory in
4D. An a priori arbitrary new coupling $\kappa$ is then needed to
reabsorb these divergences. In general, it can be made a unique
function of $g^2$ such that $W(J)$ obeys a standard homogeneous
linear renormalization group equation \cite{Dudal:2005}. This is a
good sign, as we do not want new independent couplings entering our
action or results. A nice feature of the LCO under study, i.e.
$(\lco)$, is that divergences $\propto J^2$ are in fact absent in
the correlators, so there is even no need for  the coupling $\kappa$
here. The argument goes as follows. The Ward identitites prohibit
terms in $J\gamma^2$ from occurring. Notice that this is not a
trivial point, as naively we expect it to occur from the dimensional
point of view. It is only by making use of the extended action and
its larger symmetry content that we can exclude a term $\propto
J\gamma^2$ from the game. Hence, we can set $\gamma^2=0$ to find the
vacuum divergence structure $\propto J^2$, as we will employ as
usual mass independent renormalization schemes like the
$\MSbar$ scheme. Now, there are two ways to understand that no
divergences in $J$ will occur. Firstly, at the level of the action
is easily recognized that the term $g\left(
\partial _{\nu } \overline{\omega }_{i}^{a}\right) f^{abm}\left(
D_{\nu }c\right)^{b}\varphi _{i}^{m}$ in the action is irrelevant
for the computation of the generating functional as the associated
vertices cannot couple to anything without external $\omega$- and
$c$-legs. Thus forgetting about this term, the
$(\overline{\varphi},\varphi)$- and
$(\overline{\omega},\omega)$-integrations can be done exactly, and
they neatly cancel due to the opposite statistics of both sets of
fields. Hence, all $J$-dependence is in fact lost, and a fortiori no
divergences arise. Secondly, for $\gamma^2=0$, the action $S^{\prime\prime}_{\gamma^2 = 0}$ is BRST invariant, $sS^{\prime\prime}_{\gamma^2 = 0}=0$. Consequently, the vacuum correlators $\Braket{\bigl(\overline{\varphi}\varphi-\overline{\omega}\omega\bigr)_x\bigl(\overline{\varphi}\varphi-\overline{\omega}\omega\bigr)_y}=
\Braket{s\left[(\varphi\overline{\omega})_x\bigl(\overline{\varphi}\varphi-\overline{\omega}\omega\bigr)_y\right]}=0$. Therefore, we have again proven that no divergences in $J$ appear. For $\gamma^2\neq0$, the BRST transformation $s$ no longer generates
a symmetry (see section V), hence a nonvanishing result for the
correlator
$\Braket{\bigl(\overline{\varphi}\varphi-\overline{\omega}\omega\bigr)_x\bigl(\overline{\varphi}\varphi-\overline{\omega}\omega\bigr)_y}$
or the condensate
$\Braket{\overline{\varphi}\varphi-\overline{\omega}\omega}$ is allowed. A nonvanishing VEV for our new mass operator is thus exactly allowed since the BRST is already broken by the restriction to the horizon. From the first viewpoint, the $(\overline{\varphi},\varphi)$- and $(\overline{\omega},\omega)$-integrations will no longer cancel against each other, giving room for $J$-dependent contributions in the generating functional, albeit without generating any new
divergences.

\subsection{Modifying the effective action in order to stay within the horizon}
\subsubsection{Extended action}
A very important fact is to check if it is still possible to stay within the
  Gribov region $\Omega$, after adding this new mass term.
 This can be investigated with the help of the ghost propagator $\mathcal{G}(k^2)$,   which
 can be easily read off from the Feynman diagrams depicted in FIG.~1,
\begin{figure}[t]
  \centering
      \includegraphics[width=12cm]{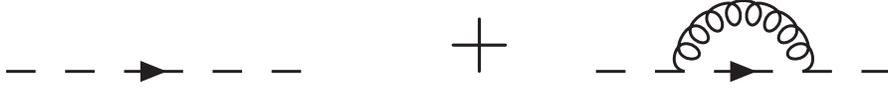}
  \label{ghostfigure}
  \caption{The one loop corrected ghost propagator.}
\end{figure}
\begin{eqnarray} \label{ghostpropagator1}
\mathcal{G}^{ab}(k^2) &=&  \delta^{ab} \mathcal{G}(k^2) ~=~ \delta^{ab}\left( \frac{1}{k^2} +
\frac{1}{k^2} \left[g^2 \frac{N}{N^2 - 1} \int \frac{\d^4
q}{(2\pi)^4} \frac{(k-q)_{\mu} k_{\nu}}{(k-q)^2}
 \Braket{A^a_{\mu}A^a_{\nu}}\right] \frac{1}{k^2} \right) + \mathcal{O}(g^4) \nonumber\\
&=& \delta^{ab} \frac{1}{k^2} (1+ \sigma(k^2)) +
\mathcal{O}(g^4)\;,
\end{eqnarray}
with
\begin{eqnarray}\label{ghi}
\sigma(k^2) &=& \frac{N}{N^2 - 1} \frac{g^2}{k^2}\int \frac{\d^4 q}{(2\pi)^4} \frac{(k-q)_{\mu} k_{\nu}}{(k-q)^2} \Braket{A^a_{\mu}A^a_{\nu}} \;.
\end{eqnarray}
Going back to the original formulation of Gribov
\cite{Gribov:1977wm}, being inside the   region
$\Omega$, is equivalent  to state that
\begin{eqnarray}
\sigma(k^2) \leq 1 \;,
\end{eqnarray}
which is called the \textit{no-pole condition}. In this case, the ghost propagator can be rewritten in the following form,
\begin{eqnarray}\label{ghostom}
\mathcal{G}(k^2) &=&  \frac{1}{k^2} \frac{1}{1 - \sigma(k^2)} + \mathcal{O}(g^4)\;,
\end{eqnarray}
which represents the fact that we are working at the level of the
inverse propagator or equivalently, at the level of the 1PI
$n$-point functions, which are generated by the effective action
$\Gamma$. This form is more natural, as we can now impose the gap
equation \eqref{gapgamma}, which is also formulated at the level
of the effective action. However, in the next section, it shall
become clear that the current action $S''$ does not guarantee us
that we are located within  the region $\Omega$ as
$\sigma(0) \geq 1$. Therefore, we add a second term to the action,
$S_{\en}$, given by
\begin{eqnarray}\label{nieuweterm}
S_{\en} &=& 2 \frac{d (N^2 -1)}{\sqrt{2 g^2 N}}  \int \d^d x\ \varsigma \ \gamma^2 J \;
\end{eqnarray}
with $\varsigma$ a new parameter. We have introduced the
particular prefactor of $ 2 \frac{d (N^2 -1)}{\sqrt{2 g^2 N}}$ for
later convenience. As it is a constant term, is it comparable with
the term $\left( - \int \d^4 x 4 (N^2 - 1) \gamma^4 \right)$ in
the original Gribov-Zwanziger formulation \eqref{7}. Therefore, it
can be responsible for allowing us to stay inside the Gribov
horizon by  enabling $\sigma$ to be smaller than 1. The
explicit calculation of $\sigma$ will be done in the next section,
but we can already intuitively sketch the reasoning why $\sigma$
will be altered. As this new term is independent of the fields, it
will only enter the expression for the vacuum energy. However, due
to the gap equation \eqref{gapgamma}, it will also enter in the
expression of the ghost propagator (and analogously any other
quantity which contains $\gamma^2$). Recapitulating, the complete
action now reads,
\begin{eqnarray}
S''' &=& S'' + S_{\en}
\end{eqnarray}
with $S''$ given in equation \eqref{extended_action}.

\subsubsection{Renormalizability}
The renormalizability of $S'''$ can be easily verified. Therefore, we replace $S_{\en}$ with
\begin{eqnarray}
\Sigma_{\en} &=&  \int \d^4 x \varsigma \Theta J \;,
\end{eqnarray}
with $\Theta$ a color singlet and BRST invariant source, $s \Theta =0$. In the end, we give $\Theta$ the physical value of
\begin{eqnarray}
\left. \Theta \right|_{\phys} &=&  2 \frac{d (N^2 -1)}{\sqrt{2 g^2 N}} \gamma^2 \;,
\end{eqnarray}
to return to the original action $S'''$. Again, we embed the action $S'''$ into a larger action $\Sigma''$,
\begin{eqnarray}
\Sigma '' &=& \Sigma' + \Sigma_{\en} \;,
\end{eqnarray}
with $\Sigma'$ given by \eqref{61}. Firstly, as it is easily checked, the term $\Sigma_{\en}$ can only give rise to an additional harmless classical breaking in the Ward identities. Therefore, all the previous Ward identities will remain valid. Secondly, we have the following additional Ward identity,
\begin{eqnarray}
\frac{\delta \Sigma '' }{\delta \Theta} &=&  \varsigma  J\;.
\end{eqnarray}
which implies that the counterterm is independent from $\Theta$. Taking these two argument together, we can conclude that the counterterm will be exactly the same as before, given by \eqref{tegenterm}. Therefore,
\begin{eqnarray}
\frac{\varsigma_0 \gamma_0^2 J_0}{ g_0}   &=& \frac{\varsigma \gamma^2 J}{ g}
\end{eqnarray}
and consequently, no new renormalization factor is necessary,
\begin{eqnarray}
Z_{\varsigma} &=& Z_g Z_{\gamma^2}^{-1} Z_J^{-1} \;.
\end{eqnarray}

\subsubsection{Boundary condition}
Introducing a new parameter $\varsigma$, requires a second gap
equation  in order to determine this new parameter. We
recall that, in the case  in which $M^2 = 0$ or
equivalently in the original Gribov-Zwanziger formulation,
 we have
\begin{eqnarray}
\sigma(k^2 \approx 0) &=& 1 - C k^2 \;,
\end{eqnarray}
with $C$ a certain positive constant, which causes the enhancement of the ghost propagator $\mathcal{G}(k^2)$ at zero momentum,
\begin{eqnarray}
\mathcal{G}(k^2 \approx 0)&\sim& \frac{1}{C k^4} \;.
\end{eqnarray}
Therefore, we know that at zero momentum, slowly switching off $M^2$, will cause $\sigma (k^2 =0)$ going to 1. It is therefore very natural to demand that this transition has to occur smoothly by imposing the following boundary condition,
\begin{eqnarray} \label{bound}
\left.\frac{\partial \sigma(0)}{\partial M^2} \right|_{M^2=0}&=& 0\;.
\end{eqnarray}

In summary, we have now two gap equations. Firstly, the gap equation $\frac{\p \Gamma}{\p \gamma^2} =0$ fixes $\gamma^2$ as a function of $M^2$ and secondly, demanding that $\left.\frac{\partial \sigma(0)}{\partial M^2} \right|_{M^2=0}= 0$ will uniquely fix $\varsigma$. This leaves us with one free parameter, $M^2$, the fixation of which shall be discussed in section \ref{sectie4}.

\section{The modified gluon and ghost propagator}
Now that we have constructed the action $S'''$, by adding two
additional terms $S_{\overline{\varphi} \varphi}$ and $S_{\en}$ to
the original Gribov-Zwanziger action, we investigate the gluon
and the ghost propagator in detail. For the calculations, we have
replaced the sources $J$ and $\tau$ with the more conventional
mass notations $M^2$ resp.~$m^2$.

\subsection{The gluon propagator}
We shall first examine the tree level gluon propagator. In order to calculate this free gluon propagator we only
need that part of the free action $S^{\prime \prime}$ containing the $A$-fields and
the $\varphi$, $\overline{\varphi}$-fields. This free action reads
\begin{eqnarray}
    S^{\prime\prime}_0 &=& \;\int \d^4x \; \left[ \frac{1}{4} \left( \p_{\mu} A_{\nu}^a - \p_{\nu}A_{\mu}^a
\right)^2 + \frac{1}{2\alpha} \left( \p_{\mu} A^a_{\mu} \right)^2 +
\overline{\varphi}^{ab}_{\mu} \p^2 \varphi^{ab}_{\mu} - \gamma^2
g(f^{abc}A^a_{\mu} \varphi_{\mu}^{bc} + f^{abc}A^a_{\mu}
\overline{\varphi}^{bc}_{\mu} ) - M^2
\overline{\varphi}_{\mu}^{ab}\varphi_{\mu}^{ab} + \frac{m^2}{2}
A^2_{\mu} +\ldots \right]
\end{eqnarray}
where the limit $\alpha \rightarrow 0$ is understood in order to
 recover the Landau gauge. The ``$\ldots$'' stand for
the constant terms $-d (N^2 -1) \gamma^4$ and $ 2 \frac{4 (N^2
-1)}{\sqrt{2 g^2 N}}  \varsigma \ \gamma^2 M^2 $ and other terms
in the ghost- and $\omega, \overline{\omega}$-fields irrelevant
for the calculation of the gluon propagator. Next, we integrate
out the $\varphi$- and $\overline{\varphi}$-fields. As we are only
interested in the gluon propagator, we simply use the equations of
motion, $\frac{\p S^{\prime\prime}_0 }{\p
\overline{\varphi}_{\mu}^{bc}} = 0$ and $\frac{\p
S^{\prime\prime}_0 }{\p\varphi_{\mu}^{bc}} = 0$, which
give
\begin{eqnarray}
    \varphi^{bc}_{\mu} = \overline{\varphi}^{bc}_{\mu} =  \frac{1}{\p^2 - M^2}
\gamma^2 g f^{abc} A^a_{\mu} \;.
\end{eqnarray}
We use this result to rewrite $ S^{\prime\prime}_0$,
\begin{eqnarray}
 S^{\prime\prime}_0&=& \; \int \d^4 x \; \left[ \frac{1}{4} \left(
\p_{\mu} A_{\nu}^a - \p_{\nu}A_{\mu}^a \right)^2+
\frac{1}{2\alpha} \left(\p_{\mu} A^a_{\mu} \right)^2 +
\frac{m^2}{2}A^2_{\mu} + \gamma^4 g^2 f^{abc} A_{\mu}^a
\frac{1}{\p^2 -M^2} f^{dbc} A^d_{\mu} - 2 \gamma^4
g(f^{abc}A^a_{\mu} \frac{1}{\p^2 - M^2} g f^{dbc} A_{\mu}^d )
+ \ldots \right] \nonumber\\
&=&  \; \int \d^4x \; \left[ \frac{1}{4} \left( \p_{\mu} A_{\nu}^a -
\p_{\nu}A_{\mu}^a \right)^2+ \frac{1}{2\alpha} \left( \p_{\mu}
A^a_{\mu} \right)^2 + \frac{m^2}{2}A^2_{\mu} - N \gamma^4 g^2
A_{\mu}^a \frac{1}{\p^2 -M^2} A^a_{\mu} + \ldots \right]\;.
\end{eqnarray}
The last step is explained with the following relation,
\begin{eqnarray}
f^{abc}f^{dbc} &=& N \delta^{ad} \;,
\end{eqnarray}
and we restrict ourselves to the color group $SU(N)$ throughout. We
continue rewriting $ S^{\prime\prime}_0$ so we can easily read the
gluon propagator
\begin{eqnarray}
 S^{\prime\prime}_0&=&  \; \int \d^4 x \; \left[ \frac{1}{2} A^a_{\mu} \Delta^{ab}_{\mu\nu} A^b_{\nu} + \ldots \right] \;, \nonumber\\
\Delta^{ab}_{\mu\nu} &=&\left[ \left(-\p^2 + m^2 - \frac{2 g^2 N
\gamma^4}{\p^2 - M^2} \right) \delta_{\mu\nu} - \p_{\mu}\p_{\nu}
\left(\frac{1}{\alpha} - 1\right) \right] \delta^{ab} \;.
\end{eqnarray}
The gluon propagator can be determined by taking the inverse of
$\Delta^{ab}_{\mu\nu}$ and converting it to momentum space. Doing
so, we find the following expression
\begin{eqnarray}
  \; \Braket{ A^a_{\mu}(p) A^b_{\nu}(-p)} &=& \frac{1}{p^2+ m^2 + \frac{2g^2 N \gamma^4}{p^2 + M^2}}\left[\delta_{\mu\nu} - \frac{p_{\mu}p_{\nu}}{p^2} \right]\delta^{ab} \nonumber\\
 &=& \underbrace{\frac{p^2 + M^2}{p^4 + (M^2+m^2)p^2 + 2 g^2 N \gamma^4 + M^2 m^2 }}_{\mathcal{D}(p^2)}\left[\delta_{\mu\nu} - \frac{p_{\mu}p_{\nu}}{p^2} \right]\delta^{ab} \;. \label{gluonprop}
\end{eqnarray}
From this expression we can already make two observations:
\begin{itemize}
\item $\mathcal{D}(p^2)$ enjoys infrared suppression.
\item
$\mathcal{D}(0)\propto M^2$, so the gluon propagator does not vanish
at the origin. Even if we set $m^2=0$ we still find a nonvanishing
gluon propagator, so we want to stress that this different result is
clearly due to the novel mass term proportional to $\overline{\varphi}\varphi-\overline{\omega}\omega$.
\end{itemize}
In section \ref{var} we shall uncover a third property, namely that
$\mathcal{D}(p^2)$ displays a positivity violation. Also this
observation is in accordance with the latest lattice results
\cite{Bowman:2007du}.

\subsection{The ghost propagator}
The observation that $m^2 =0$ does not qualitatively alter the
gluon propagator, will be repeated for the ghost propagator.
Henceforth, we set $m^2 =0$, which also improves the
readability of the paper. However, all calculations
could in principle be repeated with the inclusion of the mass $m^2$.  \\

We start with the expression for the ghost propagator. Substituting the expression of the gluon propagator, we find,
\begin{eqnarray}\label{exact}
\sigma(k^2) &=& \frac{N}{N^2 - 1} \frac{g^2}{k^2}\int \frac{\d^d q}{(2\pi)^d} \frac{(k-q)_{\mu} k_{\nu}}{(k-q)^2} \Braket{A^a_{\mu}A^a_{\nu}} \nonumber\\
&=& Ng^2 \frac{k_{\mu} k_{\nu}}{k^2} \int \frac{\d^d q}{(2\pi)^d} \frac{1}{(k-q)^2} \frac{q^2 + M^2}{q^4 + M^2 q^2 + \lambda^4 }\left[ \delta_{\mu\nu} - \frac{q_{\mu}q_{\nu}}{q^2} \right]
\end{eqnarray}
where we have also used equation \eqref{lambda4}. As we are interested in the infrared behavior of this propagator, we expand the previous expression for small $k^2$,
\begin{eqnarray}\label{sigmaex}
\sigma (k^2\approx 0) &=&  Ng^2 \frac{k_{\mu} k_{\nu}}{k^2} \frac{d-1}{d} \delta_{\mu\nu} \int \frac{\d^d q}{(2\pi)^d} \frac{1}{q^2} \frac{q^2 + M^2}{q^4 + M^2 q^2 + \lambda^4 } + O(k^2)\nonumber\\
&=&  Ng^2 \frac{d-1}{d}  \int \frac{\d^d q}{(2\pi)^d} \frac{1}{q^2} \frac{q^2 + M^2}{q^4 + M^2 q^2 + \lambda^4 } + O(k^2) \;.
\end{eqnarray}
For later use, let us rewrite $\sigma(0)$ as
\begin{eqnarray}\label{3}
\sigma (0) &=&   Ng^2 \frac{d-1}{d}  \int \frac{\d^d q}{(2\pi)^d} \frac{1}{q^4 + M^2 q^2 + \lambda^4 } +  Ng^2 M^2 \frac{d-1}{d}\int \frac{\d^d q}{(2\pi)^d} \frac{1}{q^2} \frac{ 1}{q^4 + M^2 q^2 + \lambda^4 } \;.
\end{eqnarray}
Notice that the first integral in the right hand side of equation \eqref{3} diverges while the second integral is UV finite in $4D$.\\

We continue with the derivation of the gap equations as we would like to write $\lambda^2$ as a function of $M^2$, i.e. $\lambda^2(M^2)$, in expression \eqref{3}. Firstly, we calculate the horizon condition \eqref{gapgamma} explicitly starting from the effective action. The one loop effective action $\Gamma_\gamma^{(1)}$ is obtained from the quadratic part of our action $S^{\prime\prime}$
\begin{equation}
\e^{-\Gamma_\gamma ^{(1)} }=\int \d\Phi \e^{-S^{\prime\prime}_0}\;,
\end{equation}
This time, the terms $-d(N^2 - 1) \gamma^4$ and $ 2\frac{d (N^2 -1)}{\sqrt{2 g^2 N}}  \varsigma \ \gamma^2 M^2 $ have to be maintained, as they will enter the horizon condition. After a straightforward calculation the one loop effective action in $d$ dimensions yields,
\begin{eqnarray}
\Gamma_\gamma^{(1)} &=& -d(N^{2}-1)\gamma^{4} + 2\frac{d (N^2 -1)}{\sqrt{2 g^2 N}}  \varsigma \ \gamma^2 M^2 +\frac{(N^{2}-1)}{2}\left( d-1\right) \int \frac{\d^{d}q}{\left(
2\pi \right) ^{d}} \ln \frac{q^4 + M^2 q^2 + 2 g^2 N \gamma^2}{ q^2 + M^2} \;.
\end{eqnarray}
Setting $\lambda^4 = 2 g^2 N \gamma^4$ (see equation \eqref{lambda4}), we rewrite the previous expression,
\begin{eqnarray}
\mathcal{E}^{(1)} &=&  \frac{\Gamma_\gamma^{(1)}}{N^2 - 1} \frac{2 g^2 N}{d} ~=~ - \lambda^4  + 2 \varsigma \lambda^2 M^2 + g^2 N \frac{ d-1}{d} \int \frac{\d^{d}q}{\left(
2\pi \right) ^{d}} \ln \frac{q^4 + M^2 q^2 + \lambda^4}{ q^2 + M^2} \;,
\end{eqnarray}
and apply the gap equation \eqref{gapgamma},
\begin{eqnarray}\label{gapeq}
\frac{\p \mathcal{E}^{(1)}  }{\p \lambda^2} &=& 2 \lambda^2\left( -1 + \varsigma \frac{M^2}{\lambda^2} + g^2 N \frac{d-1}{d} \int \frac{\d^{d}q}{\left(
2\pi \right) ^{d}}  \frac{1}{q^4 + M^2 q^2 + \lambda^4} \right) ~=~ 0\;.
\end{eqnarray}
Secondly, we impose the boundary condition \eqref{bound} in order to obtain an explicit value for $\varsigma$. Instead of explicitly starting from expression \eqref{exact} to fix $\varsigma$, there is a much simpler way to find the corresponding $\varsigma$. Therefore, we act with $\frac{\p}{\p M^2}$ on the gap equation \eqref{gapeq}. Subsequently setting $M^2 =0$, gives
\begin{equation}\label{simpel}
 \varsigma\frac{1}{\lambda^2(0)}    -\frac{d-1}{d}g^2N\int    \frac{\d^dq}{(2\pi)^d}\frac{1}{q^2}\frac{1}{q^4+\lambda^4(0)}~=~0\;,
\end{equation}
where we imposed \eqref{bound}.  Proceeding, we find
\begin{eqnarray}\label{laatste}
    &&-\frac{d-1}{d}g^2N\int  \frac{\d^dq}{(2\pi)^d}\frac{1}{q^2}\frac{1}{q^4+\lambda^4(0)}+\varsigma\frac{1}{\lambda^2(0)}~=~0 \nonumber\\
    \Rightarrow&&  \varsigma ~=~ \lambda^2(0) \frac{3}{4}g^2N\int  \frac{\d^4q}{(2\pi)^4}\frac{1}{q^2}\frac{1}{q^4+\lambda^4(0)}  \nonumber\\
        \Rightarrow&& \varsigma ~=~  \frac{3g^2 N}{128 \pi}\;,
\end{eqnarray}
which determines $\varsigma$ at the current order.\\

With the help of the latter two gap equations \eqref{gapeq} and \eqref{laatste}, we can rephrase the correction to the self energy of the ghost. Combining equation \eqref{3} and \eqref{gapeq} we can write
\begin{eqnarray}\label{sigmapre}
\sigma(0) &=& 1 + M^2 g^2 N \frac{d-1}{d}\int \frac{\d^d q}{(2\pi)^d} \frac{1}{q^2} \frac{ 1}{q^4 + M^2 q^2 + \lambda^4(M^2) } - \varsigma \frac{M^2}{\lambda^2(M^2)}\;.
\end{eqnarray}
From this expression, we can make several observations. Firstly, when $M^2 =0$, from the previous expression it immediately follows that
\begin{eqnarray}
\sigma(0) &=& 1\;,
\end{eqnarray}
which  gives back the ordinary Gribov-Zwanziger result \cite{Gribov:1977wm,Zwanziger:1989mf,Zwanziger:2001kw,Dudal:2005}. Indeed, from the previous expression, one derived that the ghost propagator,
\begin{eqnarray*}
  \mathcal{G}(k^2) &=& \frac{1}{k^2} \frac{1}{1 - \sigma(k^2)} \;,
\end{eqnarray*}
is enhanced and behaves like $1/k^4$, for $k^2 \approx 0$. Secondly, when $M^2 \not= 0$, we notice that the ghost propagator is no longer enhanced and behaves like $1/k^2$ as already found in \cite{Dudal:2007cw}, which is in qualitative agreement with the latest lattice results. This behavior is clearly due to the novel mass
term $ M^2\int \d^4 x\;\left( \overline{\varphi}^a_i \varphi^a_{i}
- \overline{\omega}^a_i \omega^a_i \right)$. Thirdly, we see that
the term in $\varsigma$ is crucial in
order to obtain a $\sigma(0)$ which is smaller than 1. Omitting
this term would result in $\sigma(0) > 1$ in the case that $M^2
\not= 0$. However, including this term, we can easily prove that $
\sigma \leq 1$. Indeed, taking expression \eqref{sigmapre} and
replacing $\varsigma$ with the integral in \eqref{laatste}, we
find
\begin{eqnarray}
\sigma(0) &=& 1 + M^2 g^2 N \frac{3}{4}\int \frac{\d^4 q}{(2\pi)^4} \frac{1}{q^2} \frac{ 1}{q^4 + M^2 q^2 + \lambda^4(M^2) } -  \frac{M^2}{\lambda^2(M^2)} \lambda^2(0) \frac{3}{4}g^2N\int  \frac{\d^4q}{(2\pi)^4}\frac{1}{q^2}\frac{1}{q^4+\lambda^4(0)} \nonumber\\
&=& 1 + \frac{3}{4} \frac{M^2}{\lambda^2(M^2)} g^2 N \int \frac{\d^4 p}{(2\pi)^4} \frac{1}{p^2} \frac{ 1}{p^4 + \frac{M^2}{\lambda^2(M^2)} p^2 + 1 } -   \frac{3}{4} \frac{M^2}{\lambda^2(M^2)} g^2 N \frac{3}{4}g^2N\int  \frac{\d^4 p}{(2\pi)^4}\frac{1}{p^2}\frac{1}{p^4+1} \nonumber\\
&=& 1 - \frac{3 x^2}{4}  g^2 N \int \frac{\d^4 p}{(2\pi)^4} \frac{1}{p^2}  \frac{ 1}{(p^4 + x p^2 + 1)(p^4 + 1) }  \;,
\end{eqnarray}
with $x = \frac{M^2}{\lambda^2(M^2)} \geq 0$, hence  $\sigma(0)
\leq 1 $. At this point, we can really appreciate the role of the
novel vacuum term \eqref{nieuweterm}. It serves as a stabilizing
term for the horizon condition.  Indeed, without the
term \eqref{nieuweterm}, we would end up outside of the Gribov
region for some $k^2>0$, even for an infinitesimal\footnote{Notice
that we must take $M^2 \geq 0$ to avoid unwanted tachyonic
instabilities. } $M^2>0$. In this sense, the action $S'''$
constitutes a refinement of the original Gribov-Zwanziger action,
which is a smooth limiting case of $S'''$.\\

For later use, we can evaluate the integral in expression \eqref{sigmapre} as it is finite. The explicit one loop value for $\sigma(0)$ yields
\begin{eqnarray}\label{sigmafinal}
\sigma(0) &=& 1 + M^2 \frac{3 g^2 N}{64 \pi^2}  \frac{1}{ \sqrt{M^4 - 4 \lambda^4}} \left[  \ln \left(  M^2  +  \sqrt{M^4 - 4 \lambda^4} \right) - \ln \left(M^2 -  \sqrt{M^4 - 4 \lambda^4} \right) \right] - \left( \frac{3g^2 N}{128 \pi} \right) \frac{M^2}{\lambda^2(M^2)}\;,
\end{eqnarray}
where we have substituted the value \eqref{laatste} for $\varsigma$.\\

In summary, we have found a ghost propagator which is no longer
enhanced. So far, we have fixed $\lambda^2$ in function of $M^2$
and we have found a constant value for $\varsigma$. However, we
have not  yet fixed $M^2$. This will be the
task of the next section.

\section{A dynamical value for $M^2$ \label{sectie4}}
 Up to this point, we have only introduced the mass $M^2$ by hand, however it is recommendable to obtain a dynamical value for this parameter. We shall present two methods to find such a value. Firstly, we explain how to obtain a dynamical value for $M^2$ with the help of the effective action. However, as the calculations become too involved, we investigate a second method, the variational principle, and apply this to the ghost and gluon propagator, with more success.

\subsection{The effective action and the gap equations}
We first explain the idea behind the method before going into detailed calculations. In the previous section we have derived the gluon propagator. We recall that the mass term $m^2 A^2_{\mu}$ does not qualitatively change the form of the gluon and ghost propagators, therefore we have put $m=0$ for our purpose. With $m=0$, the tree level propagator \eqref{gluonprop} yields:
\begin{eqnarray}\label{wishedform}
\mathcal{D}(p^2) &=& \frac{p^2 + M^2}{p^4 + M^2 p^2 + 2 g^2 N \gamma^4  }\;.
\end{eqnarray}
Expanding the mass $M^2$ as a series in $g^2$, gives
\begin{eqnarray}
M^2 &=& M^2_0 + g^2 M_1^2 + g^4 M_2^2 + \ldots \;.
\end{eqnarray}
We only need to consider $M_0$, which is of order unity, as we are
 considering the tree level propagator. We know that at
the end of our calculations we have to set our sources equal to
zero, or $J= M^2 =0$. If we work at lowest order, this means we
have to set $M_0 =0$ (and the gluon propagator will not display
the desired behavior). However,  going one order
higher gives:
\begin{eqnarray}
 M_0^2 + g^2 M_1^2 &=& 0     \;.
\end{eqnarray}
The last equation might imply that $M_0^2$ is no longer equal to zero, and consequently, the tree level gluon propagator will attain the desired form. Let us elaborate further on this aspect.

\subsubsection{One loop effective potential}
To implement the above-mentioned ideas, we shall first calculate the one loop energy functional. We start with the action
\eqref{extended_action}, whereby setting $m = 0$ is equivalent with putting $\tau = 0$. We replace the mass $M^2$ again with the
source $J$. In order to determine the one loop effective action, we first need the one loop energy functional $W_{0}(J)$ which we
obtain from the quadratic part of the action,
\begin{equation}
\e^{-W_{0}(J)}=\int \d\Phi \e^{-S_0^{\prime \prime \prime} }\;.
\end{equation}
From the previous expression we find for $W_{0}(J)$,
\begin{eqnarray} \label{effectiveenntuitgerekend}
W_{0}(J) &=&  - \frac{d (N^2 - 1)}{2 g^2 N} \lambda^4 +  \frac{d (N^2 -1)}{ g^2 N}  \varsigma \ \lambda^2 J + \frac{(N^{2}-1)}{2}\left( d-1\right) \int \frac{\d^{d}p}{\left(
2\pi \right) ^{d}}\ln  \left[ p^2 \left( p^{2} + \frac{\lambda^4}{p^2+ J}\right)\right]  \;.
\end{eqnarray}
We shall work in the $\MSbar$ scheme, and use a notational shorthand:
\begin{align}\label{notationalshorthand}
m_1^2 &= \frac{J - \sqrt{J^2 - 4\lambda^4  }}{2} \;, & m_2^2 &= \frac{J + \sqrt{J^2 - 4\lambda^4  }}{2} \;,
\end{align}
whereby $\lambda^2$ is defined in equation \eqref{lambda4}.
Evaluating the integrals in $W_0(J)$  gives
\begin{eqnarray} \label{effectiveenergy}
W_{0}(J) &=& - \frac{4 (N^2 - 1)}{2 g^2 N} \lambda^4 +  \frac{d (N^2 -1)}{ g^2 N}  \varsigma \ \lambda^2 M^2 +  \frac{3(N^2 - 1)}{64 \pi^2} \left( \frac{8}{3} \lambda^4 + m_1^4 \ln \frac{m_1^2}{\overline{\mu}^2}
+ m_2^4 \ln \frac{m_2^2}{\overline{\mu}^2} - J^2 \ln \frac{J}{\overline{\mu}^2}\right) \;.
\end{eqnarray}
This calculation is explained in detail in the appendix.\\

As we have determined the energy functional $W_0(J)$, we can now calculate the one loop effective action via the Legendre transform of $W(J)$. If we define
\begin{align}\label{klassiekveld}
\sigma(x) &~=~ \frac{\delta  W(J)}{\delta J(x)} & \sigma_{\mathrm{cl}} & ~=~  \frac{d (N^2 -1)}{ g^2 N}  \varsigma \ \lambda^2
\end{align}
then
\begin{eqnarray}
\widehat{\sigma}(x) &=& \sigma(x)  - \sigma_{\mathrm{cl}} =  - \frac{\int \d \Phi \left(\overline{\varphi}\varphi-\overline{\omega}\omega \right) \e^{-S'''}}{\int \d \Phi  \e^{-S'''}} \;,
\end{eqnarray}
represents the expectation value of the local composite operator, $-\left(\overline{\varphi}\varphi-\overline{\omega}\omega\right)$. The effective action is given by
\begin{eqnarray}
\Gamma(\sigma) &=& W(J) - \int \d^4 x \ J(x)\sigma(x)  \;,
\end{eqnarray}
or equivalently, as we prefer to work in the variable $\widehat{\sigma}$,
\begin{eqnarray}\label{gamma}
\Gamma(\widehat{\sigma}) &=& W(J) - \int \d^4 x \ J(x) \left( \widehat{\sigma}(x) + \sigma_{\mathrm{cl}} \right)  \;.
\end{eqnarray}
Calculating $\Gamma(\widehat{\sigma})$  by explicitly doing the inversion
is a rather cumbersome task. In  most cases one can
perform a Hubbard-Stratonovich transformation to eliminate the
term $J \left(\overline{\varphi}\varphi-\overline{\omega}\omega
\right)$ from the action and introduce a new field $\sigma'$ which
couples linearly to $J$. This  greatly simplifies the
calculation. However,  in this case, it seems
impossible to do such a transformation as a required term in $J^2$
is missing. Hence, there is no other option than to actually
perform the inversion. In order to calculate this inversion, we
shall limit ourself to constant $J$ and $\widehat{\sigma}$ as we are mainly
interested in the (space time) independent vacuum expectation
value of the operator
$-\left(\overline{\varphi}\varphi-\overline{\omega}\omega \right)$
coupled to the source $J$. This vacuum expectation value is given
by
\begin{eqnarray}
\left. \widehat{\sigma} \right|_{J=0}  &=&- \frac{\int \d
\Phi\left( \overline{\varphi}\varphi-\overline{\omega}\omega
 \right) \e^{-S}}{\int \d \Phi \e^{-S}} \;,
\end{eqnarray}
where $S$ represents the ordinary Gribov-Zwanziger action \eqref{s1eq1}. As we already have calculated $W(J)$ up to one loop is it straightforward to verify that
\begin{eqnarray}
\widehat{\sigma}~=~\frac{\p}{\p J} W_{0}(J) - \sigma_{\mathrm{cl}} &=& \frac{1}{2} \frac{3(N^2 - 1)}{64 \pi^2} J \left( 2 \ln \frac{t}{4} + \left( \sqrt{1-t} + \frac{1}{\sqrt{1-t}} \right) \ln \frac{1+ \sqrt{1-t}}{1- \sqrt{1-t}}\right) \;,
\end{eqnarray}
whereby we shortened the notation by putting $t= 4\lambda^4 / J^2 $. From the previous expression we find for the condensate
\begin{eqnarray}\label{perturbative1}
\left. \widehat{\sigma} \right|_{J=0} &=& -    \frac{3(N^2 - 1)}{64 \pi} \lambda \;.
\end{eqnarray}
This is an important result, as it indicates that a nonzero
value for the Gribov parameter $\gamma$ will result in a
nonvanishing condensate
$\Braket{-\left(\overline{\varphi}\varphi-\overline{\omega}\omega \right)}$
even at the perturbative level.
\\
We are now ready to compute the effective action up to one loop along the lines of \cite{Yokojima:1995hy}. The energy functional can be written
as a series in the coupling constant $g^2$,
\begin{eqnarray}
  W(J) &=& W_0(J) + g^2 W_1 (J) + \ldots \nonumber\\
  &=&  \sum_{i=0}^{\infty} (g^2)^i W_{i}(J).
\end{eqnarray}
As a consequence, looking at the definition \eqref{klassiekveld}, we can write
\begin{eqnarray} \label{phi}
\widehat{\sigma} &=& \widehat{\sigma}_{0}(J) + g^2 \widehat{\sigma}_{1}(J)  + \ldots \nonumber\\
&=& \sum_{i=0}^{\infty} (g^2)^i \widehat{\sigma}_{i}(J),
\end{eqnarray}
where $\widehat{\sigma}_{i}(J)$ corresponds to the $i$th order in $g^2$ (regarding $J$ as of order unity). This is called the original series. The inverted series is defined as
\begin{eqnarray} \label{J}
J &=& J_{0}(\widehat{\sigma}) + g^2 J_{1}(\widehat{\sigma})  + \ldots \nonumber\\
&=& \sum_{j=0}^{\infty} (g^2)^j J_{j}(\widehat{\sigma}),
\end{eqnarray}
with $J_{j}(\widehat{\sigma})$  the $j$th order coefficient.
Substituting \eqref{J} into \eqref{phi} gives,
\begin{eqnarray}
\widehat{\sigma} &=& \sum_{i=0}^{\infty} (g^2)^i \widehat{\sigma}_{i}\left[\sum_{j=0}^{\infty} (g^2)^j J_{j}(\widehat{\sigma})\right] \nonumber\\
&=&  \widehat{\sigma}_{0}(J_{0}(\widehat{\sigma})) + g^2 \left( \widehat{\sigma}_{0}'(J_{0}(\widehat{\sigma})) \cdot J_{1}(\widehat{\sigma}) + \widehat{\sigma}_{1}(J_{0}(\widehat{\sigma})) \right) + \ldots \;.
\end{eqnarray}
By regarding $\widehat{\sigma}$ as of the order unity and by comparing both sides of the last equation, one finds
\begin{eqnarray}
\widehat{\sigma} &=& \widehat{\sigma}_{0}\left(J_{0}(\widehat{\sigma})\right) \;,  \label{laagsteorde}\\
 J_{1}(\widehat{\sigma}) &=& - \frac{\widehat{\sigma}_{1}\left( J_{0}(\widehat{\sigma})\right)}{ \widehat{\sigma}_{0}^\prime\left( J_{0}(\widehat{\sigma})\right)} \;.\\
&\vdots& \nonumber
\end{eqnarray}
For the moment, as we are working at lowest order, we only need equation \eqref{laagsteorde}. We can invert this equation,  so we find for $J_{0}(\widehat{\sigma})$:
\begin{eqnarray}
J_{0}(\widehat{\sigma}) &=& \widehat{\sigma}_{0}^{-1}(\widehat{\sigma})\;,
\end{eqnarray}
meaning that we have to solve
\begin{eqnarray}
\widehat{\sigma} \equiv \widehat{\sigma}_0 (J_0, \lambda)&=& \frac{1}{2} \frac{3(N^2 - 1)}{64 \pi^2} J_0 \left( 2 \ln \frac{t(\lambda,J_0)}{4} + \left( \sqrt{1-t(\lambda,J_0)} + \frac{1}{\sqrt{1-t(\lambda,J_0)}} \right) \ln \frac{1+ \sqrt{1-t(\lambda,J_0)}}{1- \sqrt{1-t(\lambda,J_0)}}\right) \;,
\end{eqnarray}
for $J_0$,  so we can write
\begin{eqnarray}
J_0 &=& f(\widehat{\sigma}, \lambda)\;.
\end{eqnarray}
We immediately suspect that this inversion will not give rise to
an analytical expression. Once we have found $f(\widehat{\sigma}, \lambda)$,
we substitute this expression into the effective action,
\begin{eqnarray}\label{effectieveactie}
\Gamma(\widehat{\sigma}, \lambda) &=& W(f(\widehat{\sigma}, \lambda), \lambda) -   f(\widehat{\sigma}, \lambda) \widehat{\sigma}  \;.
\end{eqnarray}

\begin{figure}[t]
   \centering
       \includegraphics[width=8cm]{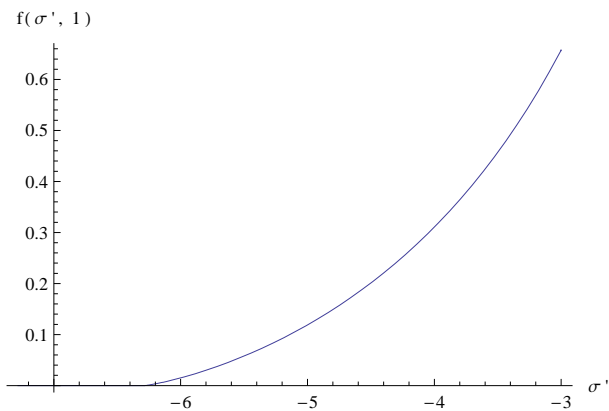}
   \caption{A plot of $f(\widehat{\sigma}' , 1)$ in terms of $\widehat{\sigma}' = \frac{2}{3} 64 \pi^2 \frac{ \widehat{\sigma}}{N^2 - 1}$ }
  \label{plot1}
\end{figure}

\begin{figure}[h]
   \centering
       \includegraphics[width=8cm]{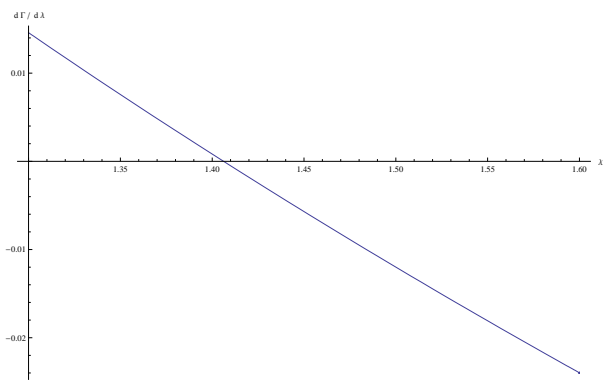}
   \caption{The horizon function $\frac{\p \Gamma}{\p \lambda}$ for $N = 3$.}
   \label{plot2}
\end{figure}

At this point, as we have found an expression for the one loop effective action, we can implement two equations to fix $\widehat{\sigma}$ and
$\lambda$. Firstly, the minimization condition reads
\begin{eqnarray}\label{gap1}
\frac{\p}{\p \widehat{\sigma}}\Gamma(\widehat{\sigma}, \lambda) &=& 0 \;,
\end{eqnarray}
and secondly, the horizon condition \eqref{gapgamma} can be translated as
\begin{eqnarray}\label{gap2}
\frac{\p}{\p \lambda}\Gamma(\widehat{\sigma}, \lambda) &=& 0 \;.
\end{eqnarray}
We start with the first gap equation.  Replacing
$\Gamma$ by equation \eqref{gamma} leads to
\begin{align}\label{gap1}
\frac{\p}{\p \widehat{\sigma}}\Gamma(\widehat{\sigma}, \lambda) &= 0 &\Rightarrow&& \frac{\p W}{\p J} \frac{\p J}{\p \widehat{\sigma}} -
\frac{\p J}{\p \widehat{\sigma}} \widehat{\sigma} - \frac{\p J}{\p \widehat{\sigma}}\sigma_{\mathrm{cl}} - J  &= 0  &\Rightarrow&&  J &=0 &\Rightarrow&& f(\widehat{\sigma}, \lambda) &=0 \;.
\end{align}
 Since there are only 2 explicit scales, $\lambda$ and $\widehat{\sigma}$, present, the first gap equation can be used to express e.g.~$\widehat{\sigma}$ in terms of $\lambda$. For the sake of a numerical
computation, we can therefore momentarily set $\lambda=1$. From FIG.~\ref{plot1} one can obtain an estimate $\widehat{\sigma}'$ of
$f(\widehat{\sigma}', 1) = 0$, with $\widehat{\sigma}' = \frac{2}{3} 64 \pi^2 \frac{ \widehat{\sigma}}{N^2 -1}\;$. Doing so, we find $\widehat{\sigma}' \approx - 6.28$, so that
\begin{eqnarray} \label{oplossing1}
\widehat{\sigma} &\approx & - 6.28  \times \left(  \frac{3(N^2 - 1)}{128 \pi^2} \right) \lambda \;,
\end{eqnarray}
which of course corresponds to the already obtained perturbative solution \eqref{perturbative1}.
The second gap equation \eqref{gap2} must
then consequently also give us back the perturbative solution. To
check this, we first calculate the perturbative result for $\lambda$ by taking the limit $J \rightarrow 0$ in expression
\eqref{effectiveenergy}
\begin{eqnarray}\label{energy}
\Gamma_0 &=&  - \frac{2 (N^2 - 1)}{ g^2 N} \lambda^4 + \frac{3(N^2 - 1)}{64 \pi^2} \left( \frac{8}{3} \lambda^4 - 2\lambda^4 \ln \frac{\lambda^2}{\overline{\mu}^2}
\right) \;.
\end{eqnarray}
Next, we take the partial derivative with respect to $\lambda$ which
gives,
\begin{eqnarray}
\frac{ \p \Gamma_0}{\p \lambda} &=&  4 \lambda^3 \left( - \frac{2 (N^2 - 1)}{ g^2 N} + \frac{3(N^2 - 1)}{64 \pi^2} \left( \frac{5}{3}  - 2 \ln \frac{\lambda^2}{\overline{\mu}^2} \right)
\right) \;.
\end{eqnarray}
The natural choice for the renormalization constant is to set $\overline{\mu} = \lambda$ to kill the logarithms. Imposing the gap equation $\frac{ \p \Gamma_0}{\p \lambda} = 0$ gives us,
\begin{eqnarray}\label{gkwadraat}
\frac{g^2 N}{16 \pi^2} &=& \frac{8}{5} \;.
\end{eqnarray}
We remark that we have neglected the solution $\gamma =0$, as explained in section \ref{GZ}. From
\begin{eqnarray}\label{gkwadraat2}
g^2(\overline{\mu}^2)&=& \frac{1}{\beta_0\ln\frac{\overline{\mu}^2}{\lms^2} }\;,\;\;\;\;\;\textrm{with }\;\;\;\;\beta_0=\frac{11}{3}\frac{N}{16\pi^2}\;,
\end{eqnarray}
and expression \eqref{gkwadraat} we find an estimate for $\lambda$:
\begin{eqnarray}\label{perturbative}
\lambda^4 &=& \e^{44/15} \;,
\end{eqnarray}
where we have worked in units $\lms = 1$. This perturbative
solution is also in compliance with \cite{Dudal:2005}. Now, we
return to the effective action \eqref{effectieveactie}. We first
take the partial derivative with respect to $\lambda$, afterwards we set
$N=3$, we explicitly replace $g^2$  by expression
\eqref{gkwadraat2} and we use the minimizing condition
\eqref{oplossing1}. Numerically, we find the following value for
$\lambda^4$:
\begin{eqnarray}
\lambda^4 &=& 1.41 \;,
\end{eqnarray}
as one can read off from FIG.~\ref{plot2}. This is exactly the perturbative result \eqref{perturbative}.
If we calculate the vacuum energy with this value for $\lambda$, we find from \eqref{energy},
\begin{eqnarray}\label{positiveenergy}
E_{\mathrm{vac}} &=& \frac{3}{64}\frac{N^2-1}{\pi^2} \e^{44/15}\;.
\end{eqnarray}
We notice that the vacuum energy is positive.

\subsubsection{Intermediate conclusion}
We can conclude at this point, that in the framework we have used, we recover only the perturbative solution. Unfortunately, at lowest order, one finds $J_0 = 0$ as explained in the beginning of this section, so we were unable to find a dynamical value for $M^2$ at first order. However, if we would be able to go one order higher, with $J_0 + g^2 J_1 = 0$, we might find $J_0 \not=0$ and consequently the gluon propagator at tree level would attain the desired form \eqref{wishedform}. In addition, we might even discover a nonperturbative solution. Unfortunately, this is not as straightforward as  at leading order. The main difficulty resides in the evaluation of two loop vacuum bubbles for the effective potential with three different mass scales.  Whilst the master integrals are known, \cite{vbv,cjj,td}, the main complication is that the propagator of \eqref{gluonprop} with $m^2$~$=$~$0$ needs to be split into standard form but this introduces the masses of \eqref{notationalshorthand} which are either complex or negative. In either scenario the master two loop vacuum bubble is known for distinct positive masses and involves several dilogarithm functions. Therefore in our case for even the simplest of mass choices the resulting dilogarithms will be complex as well as being a complicated function of $m_1^2$, $m_2^2$ and $\lambda$. Moreover, this is prior to computing the full effective potential itself by adding all the relevant combinations of master integrals together. Therefore, it seems to us that whilst such a computation could be completed in principle, currently the resulting huge expression could not possibly lend itself to a tractable analysis similar to the relatively simple one we have carried out at one loop.\\

\subsection{Applying the variational principle on the ghost propagator and the gluon propagator\label{var}}
In this section, we shall rely on variational perturbation theory in order to find a value for the hitherto arbitrary mass parameter
$M^2$. \\

Along the lines of \cite{Jackiw:1995nf}, we introduce a formal loop counting parameter $\ell\equiv 1$ by replacing the action $S$ with $\frac{1}{\ell}S$. At the same time, we replace all the fields $\Phi$ by $\sqrt{\ell}\Phi$. Symbolically,
\begin{equation}\label{vpt1}
    S(\Phi,g)\to \frac{1}{\ell}S(\sqrt{\ell}\Phi,g)\;.
\end{equation}
It is readily derived that multiplying each field with a factor of $\sqrt{\ell}$ and performing an overall $1/\ell$ rescaling is the same as replacing the coupling $g$ with
$\sqrt{\ell}g$, so we can replace \eqref{vpt1} with
\begin{equation}\label{vpt2}
    S(\Phi,g)\to S(\Phi,\sqrt{\ell}g)\;.
\end{equation}
In this fashion, the free (quadratic) part of the action is $\ell$-invariant, while every interaction terms contains powers\footnote{We recall that the perturbative expansion is one in powers of $g^2$, and thus in integer powers of $\ell$.} of $\sqrt{\ell}$. The first order in the $\ell$-expansion, obtained by setting $\ell=0$, then corresponds to the free theory. More generally, the $\ell$-expansion is equivalent with the loop expansion, where it is understood that we put the formal bookkeeping parameter $\ell=1$ at
the end.\\

The next step is to introduce the variational parameter $M^2$ into the theory. This is done in  a specific way: we add the quadratic mass term $S_M\equiv  M^2 \int  \d^4x\left[\left(\overline{\varphi}\varphi-\overline{\omega}\omega\right) + \frac{2 (N^2 -1)}{ g^2 N}   \varsigma  \lambda^2  \right]$ to the action, but substract it again at higher order in $\ell$, i.e. we consider the action
\begin{equation}\label{vpt3}
    S(\Phi,g)\to  S(\Phi,\sqrt{\ell}g)+ S_M -\ell^k S_M\;,
\end{equation}
with $k > 0$. Since $\ell\equiv1$, we did not change the actual starting action at all.\\

However, we maintain the strategy of performing an expansion in powers of $\ell$. Since the mass term is split up into 2 parts $\sim (1-\ell^k)M^2$, both parts will enter the $\ell$-expansion in a different way. At the end, we must set $\ell=1$ again. If we could compute an arbitrary quantity $\cal Q$ exactly, the $M^2$-independence would of course be apparent since the theory is not altered. However, at any finite order in $\ell$, a residual $M^2$-dependence will enter the result for $\mathcal{Q}$ due to the re-expanded powers series in $\ell$. Said otherwise, we have partially resummed the perturbative series for $\cal Q$ by making use of the parameter $\ell$. The hope is that some nontrivial information, encoded by the operator coupled to $1-\ell^k$, will emerge in the final expression for $\mathcal{Q}$. One query remains: how to handle the $M^2$ which appears in the approximate $\mathcal{Q}$? Therefore we can rely on the lore of minimal sensitivity \cite{Stevenson:1981vj}: we know that the exact $\mathcal{Q}$ cannot depend on $M^2$, hence it is very natural to demand that also at a finite order $\frac{\p \cal {Q}}{\p M^2}=0$, leading to a dynamical optimal value for the yet free parameter $M^2$. \\

The described method of variationally introducing extra parameters into
a quantum field theory provides us with a powerful tool to study nontrivial
dynamical effects in an approximate fashion, yet the calculational efforts
do not exceed those of conventional perturbation theory.\\

We still  have to choose a value for $k$. We recall
that the constant term, $S_{\en}$, was introduced in order to stay
within the horizon. Therefore, we want to retain this term when we
are applying the variational principle. However, we are working up
to first order, meaning that we shall expand the quantity
$\mathcal{Q}$ up to first order in $\ell$ and subsequently set
$\ell = 1$. Hence, taking $k =1$ is  not a good option
as the constant term would vanish and have no influence.
Therefore, a better option is to take e.g.~$k =2$, to assure the
consistency of the variational setup with the restriction to the
Gribov region. In this way, we are simply coupling the variational
parameter $M^2$ directly to the theory.

\subsubsection{The ghost propagator}
We start from the expression \eqref{ghostom} of the ghost propagator
\begin{eqnarray}
\mathcal{G}(k^2) &=&  \frac{1}{k^2} \frac{1}{1 - \sigma(k^2)}\;,
\end{eqnarray}
and apply the variational principle on the ghost propagator near zero momentum. We have,
\begin{eqnarray} \label{same}
\sigma (k^2\approx 0) &=&  Ng^2 \frac{d-1}{d}  \int \frac{\d^d q}{(2\pi)^d} \frac{1}{q^2} \frac{q^2 + M^2}{q^4 + M^2 q^2 + \lambda^4 } + O(k^2)\;.
\end{eqnarray}
As explained above, we replace $g^2 \rightarrow \ell g^2$ and $M^2 \rightarrow (1-\ell^2)M^2$.  Subsequently, we expand $\mathcal{G}(k^2)_{k^2 \approx 0}$ in powers of $\ell$ corresponding to a re-ordered loop expansion. As we have calculated the ghost propagator up to one loop, we only need to expand the above expression to the first power of $\ell$,
\begin{eqnarray}
\sigma (0) &=&  Ng^2 \ell \frac{d-1}{d}  \int \frac{\d^d q}{(2\pi)^d} \frac{1}{q^2} \frac{q^2 + M^2}{q^4 + M^2 q^2 + \lambda^4 } \;.
\end{eqnarray}
As indicated earlier, setting $\ell ~=~ 1$ gives
\begin{eqnarray}
\sigma (0) &=&  Ng^2  \frac{d-1}{d}  \int \frac{\d^d q}{(2\pi)^d} \frac{1}{q^2} \frac{q^2 + M^2}{q^4 + M^2 q^2 + \lambda^4 } \;.
\end{eqnarray}
which is exactly the same as \eqref{same}. This expression not only depends on $M^2$, but also on $\lambda^2$. However, we already know that $\lambda^2$ and $M^2$ are not independent variables, as they are related through the gap equation \eqref{gapeq},
\begin{eqnarray} \label{gapp}
  &&  -1 + \varsigma \frac{M^2}{\lambda^2} + g^2 N \frac{d-1}{d} \int \frac{\d^{d}q}{\left(
2\pi \right) ^{d}}  \frac{1}{q^4 + M^2 q^2 + \lambda^4}  ~=~ 0 \;.
\end{eqnarray}
Following the variational principle, we replace $M^2$ with $(1-\ell^2)M^2$ and $g^2$ with $\ell g^2$, expand the equation up to order $\ell^1$, and set $\ell =1$ in the end. Doing so, we recover again expression \eqref{gapp}. At this point, it can be clearly seen that $k = 1$ in equation \eqref{vpt3} would cancel the effect of the constant term $\frac{\varsigma M^2}{\lambda^2}$, while $k = 2$ is a better choice\footnote{Actually, every value for $k$, with $k \geq 2$ is allowed.}. Evaluating the integral in expression \eqref{gapp}, we find
\begin{eqnarray}
0&=&-1 + \frac{N g^2}{64 \pi^2}\left( \frac{5}{2} + 3\frac{m_1^2}{\sqrt{M^4-4\lambda^4}} \ln \frac{m_1^2}{\overline{\mu}^2} - 3\frac{m_2^2}{\sqrt{M^4-4\lambda^4}} \ln \frac{m_2^2}{\overline{\mu}^2}\right) + \varsigma \frac{M^2}{\lambda^2}
\end{eqnarray}
This integral could be similarly calculated as done in the appendix, or one could start from the effective action \eqref{effectieveactie} and derive this equation with respect to $\lambda^2$. We recall that from the boundary condition \eqref{laatste}, we have already determined $\varsigma ~=~  \frac{3g^2 N}{128 \pi}$. \\

We still require an appropriate value for $\overline{\mu}$. Therefore, we fix $\overline{\mu}^2 = \frac{3}{2}\left|M^2 + \sqrt{M^4 - 4 \lambda^4}\right|$ which was chosen as in \cite{Dudal:2005}. We have opted for this specific renormalization scale $\overline{\mu}^2$ which shall result in an acceptably small effective expansion parameter $\frac{g^2 N}{16 \pi^2}$. Consequently, from equation \eqref{gkwadraat2}, we find
\begin{eqnarray}
\frac{g^2 N}{16 \pi^2}&=& \frac{3}{11 \ln \left(\frac{3}{2}\left| M^2 + \sqrt{M^4 - 4 \lambda^4}\right| \right)}
\end{eqnarray}
in units of $\lms = 1$.\\

In summary, as $\sigma(0)$ remained the same after applying the
variational principle, we can take  the expression
\eqref{sigmafinal} for $\sigma(0)$,
\begin{eqnarray}
\sigma(0) &=& 1 + M^2 \frac{3 g^2 N}{64 \pi^2}  \frac{1}{ \sqrt{M^4 - 4 \left(\lambda^2(M^2)\right)^2}} \left[  \ln \left(  M^2  +  \sqrt{M^4 - 4 \left(\lambda^2(M^2)\right)^2} \right) - \ln \left(M^2 -  \sqrt{M^4 - 4 \left(\lambda^2(M^2)\right)^2} \right) \right] \nonumber\\
&& \hspace{11cm} - \left( \frac{3g^2 N}{128 \pi} \right) \frac{M^2}{\lambda^2(M^2)}\;.
\end{eqnarray}
where $\lambda^2(M^2)$ is determined by the gap equation,
\begin{eqnarray}\label{gapecht}
0&=&-1 + \frac{N g^2}{64 \pi^2}\left( \frac{5}{2} + 3\frac{m_1^2}{\sqrt{M^4-4\lambda^4}} \ln \frac{m_1^2}{\overline{\mu}^2} - 3\frac{m_2^2}{\sqrt{M^4-4\lambda^4}} \ln \frac{m_2^2}{\overline{\mu}^2}\right) + \frac{3g^2 N}{128 \pi} \frac{M^2}{\lambda^2}
\end{eqnarray}

\begin{figure}[t]
   \centering
       \includegraphics[width=8cm]{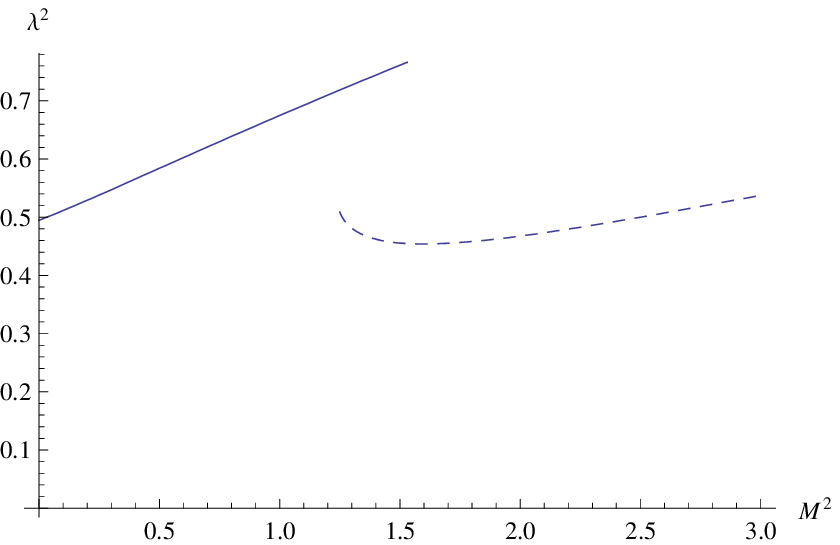}
   \caption{$\lambda^2$ in function of $M^2$ in units $\lms = 1$.}
   \label{gapdep}
\end{figure}

Before continuing the analysis, let us first have a look at the gap equation.
The gap equation solved for $\lambda^2$  as a function of $M^2$ is
depicted in FIG.~\ref{gapdep}. We find two emerging branches, displayed
by a continuous and a dashed line. The former solution exists in the
interval $[0, 1.53]$, while the latter one only exists in $[1.25, \infty[$. As
the latter branch does not exist around $M^2 =0$, we shall not consider this
solution because the boundary condition \eqref{bound} demands a smooth
transition for the $M^2 \to 0$ limit.\\

\begin{figure}[t]
   \centering
          \includegraphics[width=8cm]{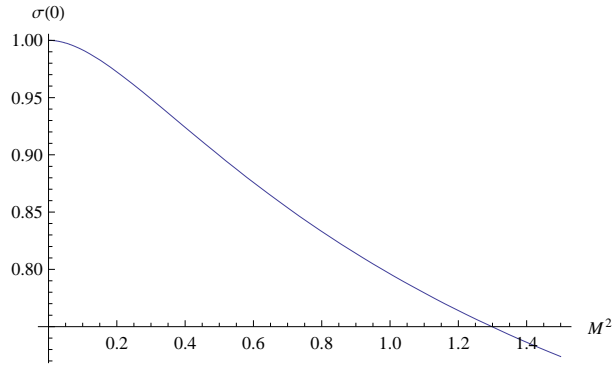}
   \caption{$\sigma(0)$ drawn in function of $M^2$ in units $\lms = 1$.}
   \label{ghost1}
\end{figure}

We can now have a closer look at the ghost propagator or equivalently $\sigma(0)$. We have graphically depicted $\sigma(0)$ in FIG.~\ref{ghost1}. Firstly, from the figure, we see that $\sigma(0)$ is nicely smaller than 1 for all $M^2$ in the interval $[0, 1.53]$. This is a remarkable fact as it implies that we have managed to stay within the horizon.  Secondly, we notice that the boundary condition $\left. \frac{\p \sigma (0)}{\p M^2} \right|_{M^2 =0} = 0$ is indeed fulfilled, which is a nice check on our result. We can now apply the minimal sensitivity approach on the quantity $\sigma(0)$. From FIG.~\ref{ghost1} we immediately see that there is no extremum. However, looking at the derivative of $\sigma(0)$ with respect to $M^2$, we do find a point of inflection at $M^2 = 0.37 \lms^2$. Demanding $\frac{\p^2 \sigma(0) }{ (\p M^2) ^2 } = 0$ is an alternative option when no extremum is found \cite{Stevenson:1981vj}. Taking this value for $M^2$, we find
\begin{eqnarray}
\sigma(0) &=&  0.93 \;.
\end{eqnarray}
The effective coupling is given by
\begin{eqnarray}\label{koppel1}
\frac{g^2 N }{ 16 \pi^2} &=& 0.53 \;,
\end{eqnarray}
which is smaller than 1.
\begin{figure}[t]
   \centering
          \includegraphics[width=8cm]{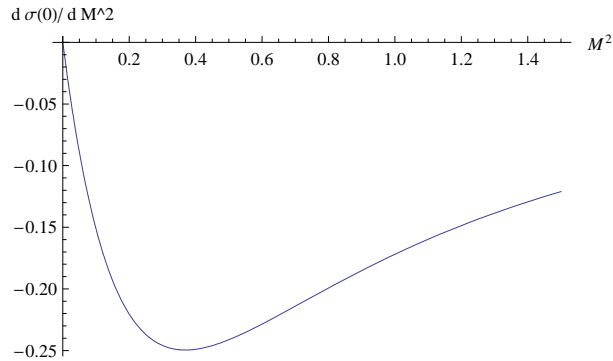}
   \caption{$ \frac{ \d \sigma(0)}{\d M^2}$ drawn in function of $M^2$ in units $\lms = 1$.}
   \label{ghost2}
\end{figure}

\subsubsection{The gluon propagator}
In order to apply the variational principle to the gluon propagator,
we require its one loop correction. Given the rather complicated
form of the propagator, obtaining the full exact expression for its
one loop correction is not possible. Indeed to appreciate how
cumbersome such an expression could be one has only to examine the
$M^2$~$=$~$m^2$~$=$~$0$ case, \cite{Gracey:2006dr}, where all the
one loop corrections to the propagators are given explicitly.
However, despite this we can still achieve our main aim of studying
the low momentum behavior of the gluon propagator corrections
directly in the zero momentum limit without knowledge of the full
correction. In \cite{Gracey:2006dr} this limit for the gluon
propagator was deduced from the exact one loop computation. However,
the resulting expression tallied with that obtained via the vacuum
bubble expansion of the underlying $2$-point functions. The latter
is a much easier technique to apply and given the equivalence of the
expressions it justifies its application to our case when
$M^2$~$\neq$~$0$. Briefly one expands the $2$-point functions
relevant to the gluon propagator construction in powers of the
external momentum $p^2$. Though the expansion is truncated at some
order such as $O((p^2)^2)$. The accompanying Feynman integrals are
massive vacuum bubbles which are essentially trivial to compute at
one loop. However, our situation is complicated significantly by the
fact that there is mixing in the quadratic part of the $\{A^a_\mu,
\varphi^{ab}_\mu\}$ sector of the tree action. Therefore in addition
to the gluon propagator, (70), we require the propagators of the
remaining fields. For this derivation here we use the conventions
and notation of the article \cite{Gracey:2006dr} for an arbitrary
color group, where the $M^2$~$=$~$0$ problem was discussed at
length. There it is evident that one has to consider the full
$\{A^a_\mu, \varphi^{ab}_\mu\}$ part of the momentum space action in
order to invert the quadratic sector to derive all the propagators.
In the Landau gauge we find the set of propagators for our situation
are
\begin{eqnarray}
\langle A^a_\mu(p) A^b_\nu(-p) \rangle &=&
\frac{\delta^{ab}(p^2+M^2)}{[(p^2)^2+M^2 p^2+C_A\gamma^4]}
P_{\mu\nu}(p)\;,
\nonumber \\
\langle A^a_\mu(p) \overline{\varphi}^{bc}_\nu(-p) \rangle &=& -~
\frac{f^{abc}\gamma^2}{\sqrt{2}[(p^2)^2+M^2 p^2+C_A\gamma^4]}
P_{\mu\nu}(p)\;,
\nonumber \\
\langle \varphi^{ab}_\mu(p) \overline{\varphi}^{cd}_\nu(-p) \rangle
&=& -~ \frac{\delta^{ac}\delta^{bd}}{(p^2+M^2)}\eta_{\mu\nu} ~+~
\frac{f^{abe}f^{cde}\gamma^4}{(p^2+M^2)[(p^2)^2+M^2
p^2+C_A\gamma^4]} P_{\mu\nu}(p)\;, \label{fullprop}
\end{eqnarray}
where the presence of $1/\sqrt{2}$ was a key ingredient in ensuring that ghost
enhancement correctly emerged in the $M^2$~$=$~$0$ case, \cite{Gracey:2006dr}.
Therefore we are confident that our extension here will include the previous
valid analysis and therefore will provide a useful check.\\

For the one loop propagator corrections one has to first compute the
corrections to all the $2$-point functions which were relevant for the
derivation of (\ref{fullprop}). From \cite{Gracey:2006dr} this is of the form
\begin{eqnarray}
&& \left(
\begin{array}{cc}
p^2 \delta^{ac} & - \gamma^2 f^{acd} \\
- \gamma^2 f^{cab} & - (p^2+M^2) \delta^{ac} \delta^{bd} \\
\end{array}
\right) + \left(\begin{array}{cc}
X \delta^{ac} & U f^{acd} \\
N f^{cab} & Q \delta^{ac} \delta^{bd} + W f^{ace} f^{bde} + R f^{abe} f^{cde}
+ S d_A^{abcd} \\
\end{array}
\right) a ~+~ O(a^2)\;, \label{twoptdef}
\end{eqnarray}
which is written with respect to the basis $\left\{\!
\frac{1}{\sqrt{2}} A^a_\mu, \varphi^{ab}_\mu \right\}$ and as we
work in the Landau gauge the common Lorentz structure,
$P_{\mu\nu}(p)$, has been factored off. The first matrix corresponds
to the tree part of the action and the quantities $X$, $U$, $N$,
$Q$, $W$, $R$ and $S$ represent the one loop corrections and we have
used the shorthand coupling constant $a$~$=$~$g^2/(16\pi^2)$. The
totally symmetric object $d_A^{abcd}$ is defined by, \cite{rsv},
\begin{equation}
d_A^{abcd} ~=~ \frac{1}{6} \mbox{Tr} \left( T_A^a T_A^{(b} T_A^c
T_A^{d)} \right)\;,
\end{equation}
where $(T_A^a)_{bc}$~$=$~$-$~$if^{abc}$ is the adjoint
representation of the color group generators. At this stage we note
that (\ref{twoptdef}) represents a formal definition and no vacuum
bubble expansion has been performed. To one loop one can formally
invert (\ref{twoptdef}) to obtain the one loop corrections to all
the propagators (\ref{fullprop}) which is
\begin{eqnarray}
&& \left(
\begin{array}{cc}
\frac{(p^2+M^2)}{[(p^2)^2+M^2p^2+C_A\gamma^4]} \delta^{cp} &
- \frac{\gamma^2}{[(p^2)^2+M^2p^2+C_A\gamma^4]} f^{cpq} \\
- \frac{\gamma^2}{[(p^2)^2+M^2p^2+C_A\gamma^4]} f^{pcd} &
- \frac{1}{(p^2+M^2)} \delta^{cp} \delta^{dq}
+ \frac{\gamma^4}{(p^2+M^2)[(p^2)^2+M^2p^2+C_A\gamma^4]} f^{cdr} f^{pqr} \\
\end{array}
\right) \nonumber \\
&&+ \left(
\begin{array}{cc}
A \delta^{cp} & C f^{cpq} \\
E f^{pcd} & G \delta^{cp} \delta^{dq} + J f^{cpe} f^{dqe} + K f^{cde} f^{pqe}
+ L d_A^{cdpq} \\
\end{array}
\right) a ~+~ O(a^2) ~.
\label{ffdef}
\end{eqnarray}
The objects $A$, $C$, $E$, $G$, $K$, $J$ and $L$ are related to the quantities
of the one loop matrix of (\ref{twoptdef}). However, as we are focussing in
this article on the gluon propagator at zero momentum then we only need the
relation for $A$ and note that the formal correction at one loop for this is
\begin{eqnarray}
A &=& -~ \frac{1}{[(p^2)^2+M^2p^2+C_A\gamma^4]^2}  \times \left[ (p^2+M^2)^2 X - C_A \gamma^2 (N+U) (p^2+M^2) + C_A \gamma^4 \left( Q + C_A R + \half C_A W \right) \right] \, .
\label{glpropdef}
\end{eqnarray}
As noted above we could in principle compute the exact form of each of the
$2$-point functions contributing to (\ref{glpropdef}) but ultimately as we will
take the $p^2$~$\rightarrow$~$0$ limit this would be unnecessarily
overcomplicated. Instead we compute those pieces of (\ref{glpropdef}) which
remain at leading order in the vacuum bubble expansion.\\

For this we need to determine the fourteen contributing Feynman
diagrams. These were generated using the {\sc Qgraf} package,
\cite{pn}, and converted into {\sc Form} input language where {\sc
Form} is a symbolic manipulation language, \cite{form}. The vacuum
bubble expansion written in {\sc Form} was applied to each integral
and expressions obtained for all the $2$-point functions. As these
depend on $M^2$ and $\gamma^2$ we were able to check that our
expressions agreed with those already determined in the
$M^2$~$=$~$0$, $\gamma^2$~$\neq$~$0$ case of \cite{Gracey:2006dr}.
Moreover, we also checked the explicit Slavnov-Taylor identities for
the renormalization of the new mass operator in the $\MSbar$ scheme
by applying the {\sc Mincer} algorithm, \cite{mincer}, written in
{\sc Form}, \cite{minform}, to the Green's function where the
operator $(\overline{\varphi}^{ab}_{\mu} \varphi^{ab}_\mu -
\overline{\omega}^{ab}_\mu \omega^{ab}_\mu)$ is inserted in an
$\omega$ $2$-point function. The resulting renormalization constants
were crucial to not only ensuring that our conventions were
consistent but also that our $2$-point function vacuum bubble
expansion is correctly finite after being fully renormalized. The
upshot of our computations is the observation that for the gluon
propagator in the zero momentum limit only $X$ is required for the
leading (momentum independent) term of (\ref{glpropdef}). Thus we
finally obtain
\begin{eqnarray}\label{oneloop}
\mathcal{D}^{(1)}(0) &=& \frac{M^2}{\lambda^4}-\frac{g^2 N}{16
\pi^2} \frac{M^4}{\lambda^8} \Biggl[  \frac{M^4}{\lambda^4}
\frac{9}{16}\sqrt{M^4 - 4 \lambda^4} \ln \frac{m_2^2}{m_1^2}
       +  \frac{M^6}{\lambda^4} \left( \frac{9}{16} \ln \frac{\lambda^4}{M^4} \right)  - \frac{15}{16}M^2 \lambda^4 \frac{1}{M^4 - 4 \lambda^4} + \frac{3}{2}\lambda^4 \frac{1}{\sqrt{M^4 - 4 \lambda^4}} \ln \frac{m_2^2}{m_1^2} \nonumber\\
      && + \frac{15}{8} \lambda^8 \frac{1}{(\!\sqrt{M^4 - 4 \lambda^4})^3} \ln \frac{m_2^2}{m_1^2}  + M^2 \left( \frac{9}{8} - \frac{21}{16} \ln \frac{\lambda^4}{M^4} \right) - \frac{3}{16}   \sqrt{M^4 - 4 \lambda^4}\ln \frac{m_2^2}{m_1^2} \Biggr]
\end{eqnarray}
for the one loop correction at zero momentum where all mass
variables correspond to renormalized ones. We note that unlike the
$M^2$~$=$~$0$ case the nonzero freezing akin to tree order is
driven by the gluon $2$-point function correction. By contrast in
the $M^2$~$=$~$0$ situation the gluon suppression at one loop
derives from the $\varphi$ $2$-point which is related to the horizon
condition and the gap equation. Also in this case $A$ will be
$O(p^2)$ and not $O(1)$ to retain suppression at one loop,
\cite{Gracey:2006dr}.\\

\begin{figure}[t]
   \centering
       \includegraphics[width=0.45\textwidth]{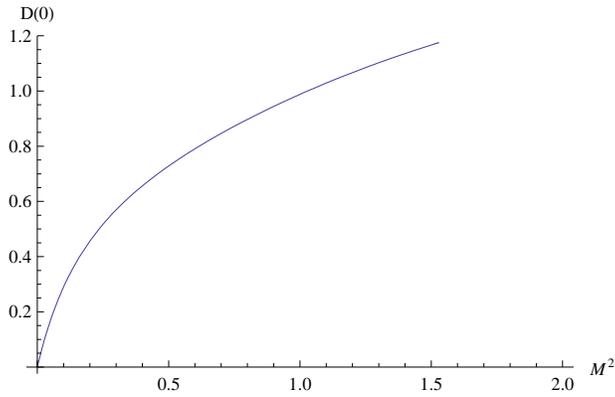}
   \caption{The gluon propagator $\mathcal{D}^{(1)}(0)$ drawn in function of $M^2$ in units $\lms = 1$.}
   \label{gluon1}
\end{figure}

We apply the variational principle to the gluon propagator in a
completely similar manner as in the case of the ghost propagator.
Therefore, we replace $M^2$ with $(1- \ell^2)M^2$ and $g^2$ with
$\ell g^2$ in the expression \eqref{oneloop}, expand up to order
$\ell^1$, and set $\ell =1$. Doing so, we find the original
expression \eqref{oneloop} for the gluon propagator back. Firstly,
we try to apply the principle of minimal sensitivity. Therefore, we
have depicted the gluon propagator in FIG.~\ref{gluon1}. First, we
notice that ${\cal D}^{(1)}(0)$ is positive for all $M^2 \in [0,
1.53]$. Unfortunately, we do find neither a minimum nor a point of
inflection in this interval. Therefore, we shall take the value of
$M^2$, which was obtained in the study of the ghost propagator (see
previous section). Hence, setting $M^2 = 0.37 \lms^2$, gives
\begin{eqnarray}
\mathcal{D}^{(1)}(0) = \frac{0.63}{\lms^2} = \frac{11.65}{ \mathrm{GeV^2}}\;.
\end{eqnarray}
Evidently, the effective coupling is still smaller than 1, cfr.~\eqref{koppel1}.\\
In summary, the infrared value of the ghost propagator and the zero
momentum gluon propagator seem to be reasonable. We find a
non-enhanced ghost propagator and a gluon propagator which is
non-zero at zero momentum. Our results for the gluon and ghost
propagator are of a qualitative nature as we are only working in a
first order approximation. In order to improve these numerical
results, higher order calculations are recommendable. This is
however far beyond the scope of the present article.

\subsubsection{The temporal correlator: violation of positivity}
With the help of the variational technique, we can also show that
the gluon propagator displays a violation of positivity. If we
rewrite the gluon propagator in the K\"{a}ll\'{e}n-Lehmann spectral
representation,
\begin{eqnarray} \label{pos}
\mathcal{D}(p^2) &=& \int_{0}^{+\infty} \d M_p^2 \frac{\rho(M_p^2)}{p^2 + M_p^2}\;,
\end{eqnarray}
$\rho(M_p^2)$ should be a positive function in order to interpret
the fields in terms of stable particles. If $\rho(M_p^2) < 0$ for
certain $M_p^2$, $\mathcal{D}(p^2)$ is positivity violating. As a
practical way to uncover this property, one defines the temporal
correlator \cite{Cucchieri:2004mf}
\begin{eqnarray}
\mathcal{C}(t)&=& \int_0^{+\infty} \d M_p \rho(M_p^2) \e^{-M_pt}  = \frac{1}{2\pi} \int_{-\infty}^{+ \infty} \e^{-ipt}\mathcal{D}(p^2) \d p\;.
\end{eqnarray}
Consequently, if we can show that $\mathcal{C}(t)$ becomes negative for certain $t$, $\rho(M_p^2)$ cannot be positive for all $M_p^2$, resulting in a positivity violating gluon propagator. If the gluon propagator vanishes at zero momentum, $\mathcal{D}(0) = 0$, one can immediately verify from \eqref{pos} that $\rho(M_p^2)$ cannot be a positive quantity. However, having $\mathcal{D}(0) \not= 0$, does not exclude a positivity violation as we shall soon find out.\\

\begin{figure}[t]
   \centering
       \includegraphics[width=8cm]{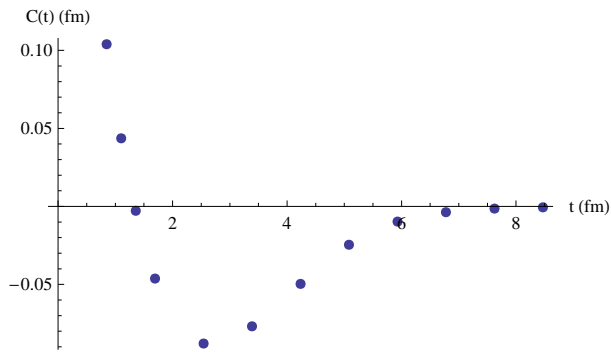}
   \caption{$\mathcal{C}(t)$ (fm) in function of $t$ (fm).}
   \label{ct}
\end{figure}

\begin{figure}[t]
   \centering
       \includegraphics[width=8cm]{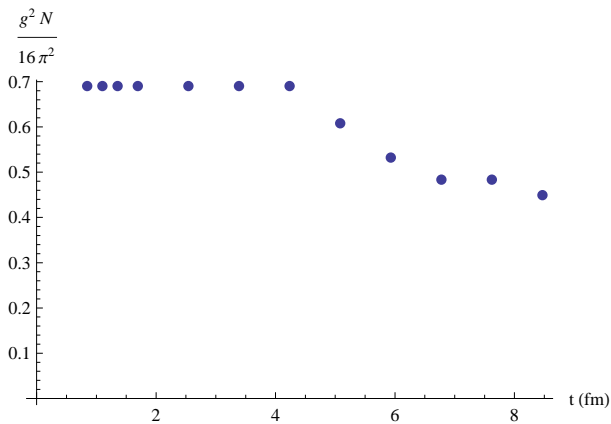}
   \caption{$g^2N/(16 \pi^2)$ in function of $t$ (fm).}
   \label{ctg}
\end{figure}

We can now apply the variational technique on the temporal correlator. At tree level, this $\mathcal{C}(t)$ is given by
\begin{eqnarray}
\mathcal{C}(t, M^2)&=& \frac{1}{2\pi} \int_{-\infty}^{+ \infty}
\e^{-ipt}\frac{p^2 + M^2}{p^4 + M^2 p^2 +
\left(\lambda^2(M^2)\right)^2} \d p\;,
\end{eqnarray}
where $\lambda^2(M^2)$ is still determined by the gap equation \eqref{gapecht}. Replacing $M^2 \rightarrow (1-\ell^2)M^2$ and $g^2 \rightarrow \ell g^2$ is redundant in this case, as we only have the
tree level gluon propagator $\mathcal{D}(p^2)$ at our disposal. We shall now implement the minimal sensitivity principle as follows: for each different value of $t$, we minimize the temporal correlator with respect to $M^2$. $C(t)$ displays a minimum at $M_{\mathrm{min}}^2 \neq 0$, for $t \gtrsim
6 /\lms $. In TABLE \ref{tabelM}, some values for $M_{\mathrm{min}}^2(t)$ for different $t$ are presented. For $t \lesssim 6$, we have taken $M^2 = 0$; it is clearly visible from the table below
that $M^2_{\min} \to 0$ for decreasing $t$.
\begin{table}[h]
\renewcommand{\arraystretch}{2}
\begin{center}
\begin{tabular}{|c||ccccc|}
        \hline
        $t $  & 6 & 7 & 8 & 9 & 10 \\
        \hline
        $M_{\mathrm{min}}^2$  & 0 & 0.16 & 0.35 & 0.51 & 0.65  \\
        \hline
\end{tabular}
\end{center}
\caption{Some $M^2_{\min}$ for different $t$ in units $\lms = 1$.}\label{tabelM}
\end{table}
The corresponding $\mathcal{C}(t,M_{\mathrm{min}}^2)$ is depicted in
\mbox{FIG. \ref{ct}}. Both the $x$-axis and $y$-axis are shown in
units fm ($1/\lms = 0.847$ fm), in order to compare our results with
\cite{Bowman:2007du, Silva:2006bs}.  Not only do we find a
positivity violating gluon propagator as $\mathcal{C}(t)$ becomes
negative, but even the shape of this function is consistent with the
lattice results\footnote{\cite{Bowman:2007du} included quarks, while
\cite{Silva:2006bs} considered gluodynamics as we are studying in
this work.} \cite{Bowman:2007du, Silva:2006bs}. Moreover, in
\cite{Bowman:2007du, Silva:2006bs}, the positivity violation starts
from $t \sim 1.5$ fm, in good agreement with our results. Finally,
FIG.~\ref{ctg} displays the corresponding values of $g^2 N/16 \pi^2$.
We can conclude that the previous  results are reliable for $t
\lesssim 8$
as $g^2N/16 \pi^2$ is smaller than one.

\subsection{A remark about the strong coupling constant}
A renormalization group invariant definition of an effective strong coupling constant $g^2_{\mathrm{eff}}$ can be written down from the knowledge of the gluon and ghost propagators as
\begin{equation}
g^2_{\textrm{eff}}(p^2)=g^2 \left(\omu^2\right) \widetilde{\mathcal{D}}\left(p^2,\omu^2\right) \widetilde{\mathcal{G}}^2 \left(p^2,\omu^2\right)\;,\label{coup}
\end{equation}
see e.g. \cite{Alkofer:2000wg}. $\widetilde{\mathcal{D}}$ and $\widetilde{\mathcal{G}}$ stand for the gluon and ghost form factor, defined by
\begin{eqnarray}
  \widetilde{\mathcal{D}}(p^2) &=& p^2\mathcal{D}(p^2)\;, \nonumber\\
  \widetilde{\mathcal{G}}(p^2) &=& p^2\mathcal{G}(p^2)\;.
\end{eqnarray}
The definition \eqref{coup} represents a kind of
nonperturbative extension of the nonrenormalization of the ghost-gluon vertex. At the perturbative level, this is assured by the Ward identity \eqref{wardid}, $Z_g=Z_c^{-1}Z_A^{-1/2}$. Usually, this is assumed to remain valid at the nonperturbative level. Although this cannot be proven, this hypothesis has been corroborated by lattice studies like \cite{Cucchieri:2006tf,Cucchieri:2008qm}.\\

In recent years, there was accumulating evidence that $g^2_{\textrm{eff}}(p^2)$ would reach an infrared fixed point different from zero: see e.g. \cite{Alkofer:2000wg, Lerche:2002ep,Zwanziger:2001kw,Pawlowski:2003hq} for a Schwinger-Dyson analysis, \cite{Gracey:2006dr,Gracey:2007bj} in the ordinary Gribov-Zwanziger approach and \cite{Bloch:2003sk,Furui:2004cx} for lattice results. These studies are mostly done in a MOM renormalization scheme, with the exception of \cite{Gracey:2006dr} where the $\MSbar$ scheme was employed.
The manifestation of this infrared fixed point was motivated in Schwinger-Dyson studies and the ordinary Gribov-Zwanziger case by means of the power law behavior of the form factors,
\begin{eqnarray}
\widetilde{\mathcal{D}}(p^2)_{p^2\approx0}&\propto&\left(p^2\right)^{2\alpha}\;,
\nonumber\\
\widetilde{\mathcal{G}}(p^2)_{p^2\approx0}&\propto&\left(p^2\right)^{-\alpha}\;,\label{form2}
\end{eqnarray}
being expressable in terms of a single exponent $\alpha$. The Schwinger-Dyson community heralded in a variety of studies the value $\alpha\approx0.595$, whereas the Gribov-Zwanziger scenario gives $\alpha=1$. Anyhow, substituting a behavior like \eqref{form2} into the definition \eqref{coup} leads to $g^2_{\textrm{eff}}(p^2)_{p^2\approx 0}\propto (p^2)^0$, opening the door for a finite value.\\

However, once again quoting the more recent large volume lattice
data of
\cite{Cucchieri:2007md,Cucchieri:2007rg,Cucchieri:2008qm,Cucchieri:2008fc},
the power law behavior \eqref{form2} seem to be excluded in favor of
\begin{eqnarray}
\widetilde{\mathcal{D}}(p^2)_{p^2\approx0}&\propto&p^2\;,
\nonumber\\
\widetilde{\mathcal{G}}(p^2)_{p^2\approx0}&\propto&\left(p^2\right)^{0}\;,\label{form3}
\end{eqnarray}
leading to a vanishing infrared effective strong coupling
constant at zero momentum since $g^2_{\textrm{eff}}(p^2)_{p^2\approx 0}\propto p^2$.
The refined analysis in this paper of the extended
Gribov-Zwanziger action, including an additional dynamical
effect, allows us to draw a similar conclusion up to the one
loop level, i.e. an infrared vanishing $g^2_{\textrm{eff}}$.
Certain lattice studies also pointed towards this particular
scenario \cite{Ilgenfritz:2006gp}.

\section{The BRST breaking in the Gribov-Zwanziger theory \label{sectie5}}
We recall here that the Gribov-Zwanziger action \eqref{s1eq1} is not
invariant under the BRST transformation \eqref{BRST}. Indeed, if we
take the BRST variation of the action \eqref{s1eq1},  one finds a
breaking term $\Delta_\gamma$ given by
\begin{eqnarray}
   \Delta_{\gamma} &\equiv& sS ~=~g \gamma^2 \int \d^4 x f^{abc} \left( A^a_{\mu} \omega^{bc}_\mu -
 \left(D_{\mu}^{am} c^m\right)\left( \overline{\varphi}^{bc}_\mu + \varphi^{bc}_{\mu}\right)  \right) \;.
 \label{deltabrst}
\end{eqnarray}
We see that the presence of the Gribov parameter $\gamma$ prevents
the action from being invariant under the BRST symmetry.
Nevertheless, this fact does not prevent the use of the
Slavnov-Taylor identity to prove the renormalizability of the
theory, which is very remarkable. Since the breaking
$\Delta_{\gamma}$ is soft, i.e.~it is of dimension two in the
fields, it can be neglected in the deep ultraviolet, where we
recover the usual notion of exact BRST invariance as well as of BRST
cohomology for defining the physical subspace
\cite{Henneaux:1992ig}. However, in the nonperturbative infrared
region, the breaking term cannot be neglected and the BRST
invariance is lost. In the following, we shall present a detailed
analysis of this breaking and of its consequences. In particular, we
shall be able to prove that the origin of this breaking can be
traced back to the properties of the Gribov region $\Omega$.
Moreover, it turns out that the existence of this breaking enables
us to give an elementary algebraic proof of the fact that the Gribov
parameter $\gamma$ is a physical parameter of the theory, entering
thus the expression of the correlation
functions of gauge invariant operators. \\

\subsection{The transversality of the gluon propagator}
The reader might wonder whether the gluon propagator still remains
transverse in the presence of the Gribov horizon. As the gluon
propagator is the connected two-point function, we ought to consider
the generator $Z^c$ of connected Green functions, which can be
constructed from the quantum effective action\footnote{This is the
generator of the 1PI Green functions.} $\Gamma$ by means of a
Legendre transformation. The renormalizability of the theory entails
that $\Gamma$ obeys the renormalized version of the Ward identity
\eqref{gaugeward}, or
\begin{equation}\label{gp1}
    \frac{\delta\Gamma}{\delta b^a}=\p_\mu A_\mu^a\;.
\end{equation}
Introducing sources $I^a(J_\mu^a)$ for the fields $b^a (A_\mu^a)$
and performing the Legendre transformation, the identity \eqref{gp1}
translates into
\begin{equation}\label{gp2}
    I^a=\p_\mu\frac{\delta Z^c}{\delta
    J_\mu^a}\;,
\end{equation}
Acting with $\frac{\delta}{\delta J_\mu^b}$ on this expression, and
by setting all sources equal to zero, we retrieve
\begin{equation}\label{gp3}
    0=\p_\mu^x\left.\frac{\delta^2 Z^c}{\delta J_\mu^a(x)
    \delta
    J_\mu^b(y)}\right|_{I,J=0}=\p_\mu^x\braket{A_\mu^a(x)A_\nu^b(y)}\;,
\end{equation}
which expresses nothing else but the transversality of the gluon
propagator.

\subsection{The BRST breaking and its consequences on the Slavnov-Taylor identity}
Let us present here a few considerations on the consequences
stemming from the BRST breaking $\Delta_{\gamma}$ appearing in the
left hand side of equation \eqref{deltabrst} of the Slavnov-Taylor
identity. Our argument will follow \cite{Capri:2007ix}. We start
from the generalized Slavnov-Taylor identity \eqref{slavnov} which
is fulfilled by the enlarged action $\Sigma$ \eqref{enlarged}. The
quantum effective action, $\Gamma = \Sigma + \hbar \Gamma^{(1)} +
\ldots$, obeys the quantum version of this Slavnov-Taylor identity
\cite{Piguet:1995er},
\begin{eqnarray}
\mathcal{S}(\Gamma ) &=&\int \d^{4}x\left( \frac{\delta \Gamma}{\delta K_{\mu }^{a}}\frac{\delta \Gamma }{\delta A_{\mu}^{a}}+\frac{\delta \Gamma }{\delta L^{a}}\frac{\delta \Gamma}{\delta c^{a}}+b^{a}\frac{\delta \Gamma}{\delta \overline{c}^{a}}+\overline{\varphi }_{i}^{a}\frac{\delta \Gamma }{\delta \overline{\omega }_{i}^{a}}+\omega _{i}^{a}\frac{\delta \Gamma }{\delta \varphi _{i}^{a}}+M_{\mu }^{ai}\frac{\delta \Gamma}{\delta U_{\mu}^{ai}}+N_{\mu }^{ai}\frac{\delta \Gamma }{\delta V_{\mu }^{ai}}\right)  = 0 \;.
\end{eqnarray}
We now pass to the Gribov-Zwanziger action, defined by
giving the sources ($M$, $N$, $U$, $V$) their physical
values \eqref{physval2} and \eqref{physval1}. As a
consequence, the physical quantum effective action
$\Gamma_{\phys}$ will now obey a broken Slavnov-Taylor identity,
\begin{eqnarray}\label{sti}
\mathcal{S}(\Gamma_{\phys} ) &=&\int \d^{4}x\left( \frac{\delta \Gamma_{\phys}}{\delta K_{\mu }^{a}}\frac{\delta \Gamma_{\phys} }{\delta A_{\mu}^{a}}+\frac{\delta \Gamma_{\phys} }{\delta L^{a}}\frac{\delta \Gamma_{\phys}}{\delta c^{a}}+b^{a}\frac{\delta \Gamma_{\phys}}{\delta \overline{c}^{a}}+\overline{\varphi }_{i}^{a}\frac{\delta \Gamma_{\phys} }{\delta \overline{\omega }_{i}^{a}}+\omega _{i}^{a}\frac{\delta \Gamma_{\phys} }{\delta \varphi _{i}^{a}}\right) \nonumber\\
&=& - \left. \int \d^4 x \left( M_{\mu }^{ai}\frac{\delta \Gamma}{\delta U_{\mu}^{ai}}+N_{\mu }^{ai}\frac{\delta \Gamma }{\delta V_{\mu }^{ai}}\right) \right|_{\phys}\nonumber\\
&=& - g \gamma^2 \left[ \int \d^4 x f^{abc} A^a_{\mu}
\omega^{bc}_\mu \cdot \Gamma_{\phys} \right] + g \gamma^2 \left[
\int \d^4 x f^{abc} \left(D_{\mu}^{am} c^m\right)\left(
\overline{\varphi}^{bc}_\mu + \varphi^{bc}_{\mu}\right) \cdot
\Gamma_{\phys} \right]\nonumber\\
&=&  -\left[ \Delta_{\gamma} \cdot \Gamma_{\phys} \right]  \;,
\label{stdelta}
\end{eqnarray}
whereby  $\left[ \Delta_{\gamma} \cdot
\Gamma_{\phys} \right]$ represents the generator of the
1PI Green functions with the insertion of the composite operator
 $\Delta_{\gamma}$.  Expression
\eqref{stdelta} generalizes at the quantum level the broken identity
of equation \eqref{deltabrst}. Once having a Slavnov-Taylor identity
like \eqref{sti} at our disposal, we can obtain relations between
different Green functions by acting on it with test operators
$\frac{\delta^n}{\delta \Psi(x_1) \ldots\delta \Psi(x_n)}$, with
$\Psi$ any field, and by setting all fields and sources equal to
zero at the end. The breaking term in the r.h.s.~of expression
\eqref{sti} will be translated into an extra contribution. In
particular, we shall obtain
\begin{eqnarray}\label{br}
\left.\frac{\delta^{n}\left[\mathcal{S}(\Gamma_{\phys}
)\right]}{\delta \Psi(x_1) \ldots\delta
\Psi(x_n)}\right|_{\mathrm{fields,sources}=0}&=&-\left.\frac{\delta^n
\left[ \Delta_{\gamma} \cdot \Gamma_{\phys} \right] }{\delta
\Psi(x_1) \ldots\delta \Psi(x_n)}\right|_{\mathrm{fields,sources}=0}
\;.
\end{eqnarray}
One sees thus that the r.h.s.~of the foregoing expression,
corresponding to
a 1PI Green function with the insertion of the composite operator $%
\Delta _{\gamma }$ and with $n$ amputated external legs of the type
$\Psi(x_1),\ldots,\Psi(x_n)$, gives precisely the modification of
the relationships among the Green functions due to the Gribov
horizon. To our understanding, the contributions stemming from the
r.h.s.~of equation (\ref{br}) should be correctly taken into account
when checking the validity of the Slavnov-Taylor identities or when
invoking Slavnov-Taylor related identities in computations when the
restriction to the Gribov horizon is
understood.\\

It is worth noticing that the breaking term of \eqref{br} will
certainly vanish if the chain $\Psi(x_1)  \ldots \Psi(x_n)$ has a
ghost number different from $+1$. Indeed, the action preserves ghost
number and the breaking term $\Delta_\gamma$ itself carries a
nonvanishing ghost charge of $+1$, so that the operator
$\frac{\delta^{n}}{\delta\Psi(x_1) \ldots \delta\Psi(x_n)}$ must
have ghost number $-1$ in order to allow for a nonvanishing contribution \eqref{br}.\\

In summary, we emphasize that the broken Slavnov-Taylor identity
\eqref{stdelta} does in fact maintain a powerful predictive
character. It allows us to establish relationships among various
Green functions of the theory in a way which takes into account the
presence of the Gribov horizon. At the same time, there exist Green
functions for which the breaking of the Slavnov-Taylor identity is
harmless. In particular, this is the case when considering gauge
invariant operators built up with only the gauge fields $A_{\mu}^a$.
For these Green functions, the physical quantum action
$\Gamma_{\phys}$ behaves as it fulfills the unbroken Slavnov-Taylor
identity, namely
\begin{eqnarray}
 \; \mathcal{S}(\Gamma_{\phys} ) &=& 0 \;.
\end{eqnarray}
The gauge invariance of the correlator $\Braket{F^2(x) F^2(y)}$
implies in fact that no useful information can be extracted for it
from the Slavnov-Taylor identities. In a loose way of speaking,
$\Braket{F^2(x) F^2(y)}$ lives on its own and is not related to
other Green functions. To formally prove this, one should add the
operator $F^2(x)$ to the action with a (BRST invariant) scalar
source $K(x)$, the (broken) Slavnov-Taylor identity \eqref{stdelta}
will remain unchanged. Hence, similarly as in the previous
subsection, by performing a Legendre transformation to pass to $Z^c$
and by acting with $\frac{\delta}{\delta K(x_i)}$ on that identity
and again setting all sources to zero, it will follow that there is
a trivially vanishing breaking term due to ghost charge
conservation.\\

\subsection{A few words on unitarity}
Certainly, the BRST breaking and its consequences on the Green
functions of the theory deserve further investigation. In this
respect one could attempt to evaluate some gauge invariant
correlation function like, for instance,  $\Braket{F^2(x) F^2(y)}$
in order to see if, despite the presence of the BRST breaking and of
a positivity violating gluon propagator, this gauge invariant
correlation function might displays a real pole in momentum space. A
first hint that something like this might happen, has been given by
a tree level computation in
\cite{Zwanziger:1989mf}.\\

As one can easily figure out, the presence of the BRST breaking
$\Delta _{\gamma }$ is related to the lack of unitarity in the gluon
sector. To our understanding, this is a manifestation of gluon
confinement: unitarity is jeopardized in the gluon sector because
gluons are confined. This is also apparent from the positivity
violation exhibited by the gluon propagator, which does not allow
for a physical interpretation of the elementary gluon excitations.
One might have the tendency to believe that the existence of the
soft breaking $\Delta _{\gamma }$ of the BRST symmetry is a welcome
feature, in particular signalling that, in a confining theory,
physics in the infrared region is not necessarily definable in the
same way as in the deep ultraviolet, where the BRST breaking could
be neglected and one recovers usual perturbation theory. As we
already stated in the beginning of this section, in the ultraviolet,
we also recover the usual notion of the BRST cohomology
\cite{Henneaux:1992ig}, allowing to prove that the ghost degrees of
freedom cancel against 2 unphysical gluon polarizations, leaving
over only 2 physical transverse polarizations, endowed with a
positive norm. In the confining regime, it is unknown what the
analogue of this scenario might be. The absence of the BRST symmetry
in the infrared does not necessarily entail that the theory is not
unitary. Certainly, the $\mathcal{S}$-matrix of the excitations of
the physical spectrum has to be unitary. But as gluons are
\emph{not} the excitations belonging to the physical spectrum,
unitarity is not to be expected in the sector described by the
elementary gluon fields. From this perspective, the question of what
the number of physical gluon polarizations might be in the
nonperturbative confining infrared sector loses its context.

\subsection{The BRST breaking as a tool to prove that the Gribov parameter is a physical parameter}
The breaking term \eqref{deltabrst} has also the interesting
consequence that it allows us to give a simple algebraic proof of
the fact that the Gribov parameter $\gamma$ is a physical parameter
of the theory, and that as such it can enter the explicit expression
of gauge invariant correlation functions like for instance
$\braket{F^{2}(x)F^{2}(y)} $ or the vacuum condensate
$\braket{F^2}$. In fact, by taking the derivative of both sides of
equation (\ref{deltabrst}) with respect to $\gamma ^{2}$ one gets,
\begin{eqnarray}
s\frac{\partial S}{\partial \gamma ^{2}}&=& \frac{1}{\gamma^2}
\Delta _{\gamma } = ~g  \int \d^4 x f^{abc} \left( A^a_{\mu}
\omega^{bc}_\mu -
 \left(D_{\mu}^{am} c^m\right)\left( \overline{\varphi}^{bc}_\mu + \varphi^{bc}_{\mu}\right)  \right)\;,  \label{d20}
\end{eqnarray}
from which, keeping in mind that the BRST operator $s$ as defined in
equation \eqref{BRST} is still nilpotent, it immediately follows
that $\frac{\partial S}{\partial \gamma ^{2}}$ cannot be cast in the
form of a BRST exact variation, namely
\begin{eqnarray}
\frac{\partial S}{\partial \gamma ^{2}}\neq s\widehat{\Delta
}_{\gamma }\;, \label{d21}
\end{eqnarray}
for some local integrated dimension two quantity $\widehat{\Delta
}_{\gamma } $. From equation (\ref{d21}) it becomes then apparent
that the Gribov parameter $\gamma ^{2}$ is a physical parameter, as
much as the gauge coupling constant $g$, for which a similar
equation holds. Furthermore, it is worth underlining that, due to
the form of the BRST operator $s$, the presence of the soft breaking
$\Delta _{\gamma }$ is, in practice, the unique way to ensure that
the Gribov parameter indeed is a physical parameter and not an
unphysical one, as it would be the case of a gauge parameter
entering the gauge fixing term. Let us suppose that the part of the
action $S_{\gamma }$ containing the Gribov parameter would be left
invariant by the BRST transformation \eqref{BRST}, namely
\begin{eqnarray}
sS_{\gamma }&=&0\;,  \label{d22}
\end{eqnarray}
instead of inducing the breaking term $\Delta _{\gamma }$. Since $S_{\gamma }$ depends on
the auxiliary fields $\left( \overline{%
\varphi }_{\mu }^{ac},\varphi _{\mu }^{ac},\overline{\omega }_{\mu
}^{ac},\omega _{\mu }^{ac}\right) $ which constitute a set of BRST
doublets\footnote{We remind here that a BRST doublet is given by a
pair $(\alpha,\beta)$ transforming as: $s\alpha=\beta$, $s\beta$ =0.
It can be shown that a BRST doublet has always vanishing cohomology,
meaning that any invariant quantity, $sF(\alpha,\beta)=0$, has
necessarily the form of an exact BRST cocycle, namely
$F(\alpha,\beta)=s{\hat F}(\alpha,\beta)$. } \cite{Piguet:1995er},
it would follow from equation (\ref{d22}) that a local integrated
polynomial $\widehat{S}_{\gamma }$ would exist such that
\begin{equation}
S_{\gamma }=s\widehat{S}_{\gamma }\;.  \label{d23}
\end{equation}
Subsequently, taking the derivative of both sides of expression
(\ref{d23}) with respect to $\gamma ^{2}$, one would obtain
\begin{equation}
\frac{\partial S_{\gamma }}{\partial \gamma ^{2}}=s\frac{\partial \widehat{S}%
_{\gamma }}{\partial \gamma ^{2}}\,,  \label{d24}
\end{equation}
a relation implying that $\gamma ^{2}$ would have the same meaning
as an unphysical gauge parameter\footnote{One easily shows that in
this case, $\frac{\p \braket{\mathcal{G}}}{\p
\gamma^2}=-\Braket{s\left(\frac{\p \widetilde{S}}{\p
\gamma^2}\mathcal{G}\right)}=0$ for any gauge invariant operator
$\mathcal{G}$.}. In turn, this would imply that correlation
functions of gauge invariant operators would be completely
independent from $\gamma^2 $. We see thus that the presence of the
soft breaking term $\Delta _{\gamma }$ plays an important role,
ensuring that $\gamma ^{2}$ is a relevant parameter of the theory.
The same conclusion also holds when the Gribov-Zwanziger action is
supplemented by the BRST invariant mass term
\eqref{extended_action}. The existence of the breaking $\Delta
_{\gamma }$ thus seems to be an important ingredient to introduce a
nonperturbative mass gap in a
local and renormalizable way.\\

A question which arises almost naturally is whether it might be
possible to modify the BRST operator, i.e. $s\rightarrow s_m$, in
such a way that the new operator $s_m$ would be still nilpotent,
while defining an exact symmetry of the action, $s_mS''=0$. Although
we are not going to give a formal proof, we can present a simple
argument discarding such a possibility. We have already observed
that the BRST transformation \eqref{BRST} defines an exact symmetry
of the action when $\gamma =0$, which corresponds to the physical
situation in which the restriction to the Gribov region has not been
implemented. Hence, it appears that one should search for possible
modifications of the BRST operator which depends on $\gamma $,
namely
\begin{eqnarray}
s_m&=&s+s_\gamma  \label{ms}\;,
\end{eqnarray}
whereby
\begin{eqnarray}
s_\gamma&=&\gamma\textrm{-dependent}\;\textrm{terms}\;,
\label{mstwee}
\end{eqnarray}
so as to guarantee a smooth limit when $\gamma$ is set to zero.
However, taking into account the fact that $\gamma $ has mass
dimension one, that all auxiliary fields $\left( \overline{\varphi
}_{\mu }^{ac},\varphi _{\mu }^{ac},\overline{\omega }_{\mu
}^{ac},\omega _{\mu }^{ac}\right)$ have dimension one too, and that
the BRST operator $s$ does not alter the dimension of the
fields\footnote{It is understood that the usual canonical dimensions
are assigned to the fields $A^{a}_{\mu}$, $b^a$, $c^a$, ${\bar c}^a$
\cite{Piguet:1995er}, as shown in TABLE \ref{tabel1}. It is apparent
that the BRST operator $s$ does not alter the dimension of the
fields.}, it does not seem possible to introduce extra
$\gamma$-dependent terms in the BRST transformation of the fields
$\left( \overline{\varphi }_{\mu }^{ac},\varphi _{\mu
}^{ac},\overline{\omega }_{\mu }^{ac},\omega _{\mu }^{ac}\right) $
while preserving locality, Lorentz covariance as well as color group
structure.

\subsection{Tracing the origin of the BRST breaking}
Having clearly seen the explicit loss of the BRST symmetry, it
would be instructive to point out more precisely where this
breaking originates from. We recall that the BRST transformation
of the gluon field $A_\mu$ is in fact constructed from the
infinitesimal gauge transformations. Indeed, for an infinitesimal
 gauge parameter $\omega^a$, the corresponding gauge
transformation is determined by
\begin{eqnarray}\label{brstbreking1}
    \delta_\omega A_\mu^a&=&D_\mu^{ab}\omega^b\;,
\end{eqnarray}
which can be compared with the BRST transformation \eqref{BRST}.
Based on this identification, we shall present our argument using
infinitesimal gauge transformations. In particular,  we
shall establish the following proposition: any infinitesimal gauge
transformation of field configurations  belonging to
the Gribov region $\Omega$, necessarily gives rise to
configurations which lie \emph{outside} of $\Omega$. We can
distinguish 2 cases.
\begin{itemize}
\item \textbf{The field $A_\mu$ is not located close to the boundary $\p
\Omega$}\\
Let us consider a gauge configuration $A_\mu$ which belongs to the
Gribov region $\Omega$ but not close to its boundary $\p\Omega$ (the
horizon), thus $\p_\mu A_\mu=0$ and $-\partial_\mu D_\mu(A)>0$.
Next, consider the field $\widetilde{A}_\mu$ obtained from $A_\mu$
through an infinitesimal gauge transformation with parameter
$\omega$,
\begin{equation}\label{brstbreking2}
    \widetilde{A}_\mu=A_\mu+D_\mu(A)\omega\;.
\end{equation}
This configuration $\widetilde{A}_\mu$ cannot belong to $\Omega$.
Suppose the  contrary, then $\p_\mu
\widetilde{A}_\mu=0=\p_\mu A_\mu$ would lead to
\begin{equation}\label{brstbreking3}
    \p_\mu D_\mu(A)\omega=0\;,
\end{equation}
in contradiction with the hypothesis that $A_\mu$ is not located on
the boundary $\p\Omega$, thus there are no zero modes $\omega$
allowing for \eqref{brstbreking3} to hold.

\item \textbf{The field $A_\mu$ is located close to the boundary
$\p \Omega$}\\In this case, we can even make a more precise
statement. If $A_\mu$ lies very close to the boundary
$\partial \Omega$, we can decompose it as
\begin{eqnarray}\label{brstbreking4}
  A_\mu &=& a_\mu + C_\mu\;,
\end{eqnarray}
with $C_\mu \in \p\Omega$, thus $C_\mu$ lies on the horizon. The
shift $a_\mu$ is a small (infinitesimal) perturbation. Obviously,
$\p_\mu C_\mu = \p_\mu a_\mu=0$. Subsequently, we find
\begin{eqnarray}\label{brstbreking5}
  \widetilde{A}_\mu &=& C_\mu+a_\mu+D_\mu(C)\omega +\ldots
\end{eqnarray}
for the gauge transformed field at lowest order in the infinitesimal
quantities $\omega$ and $a_\mu$. Since
$C_\mu\in\p\Omega$ and by identifying $\omega$ with the zero mode
corresponding to $C_\mu$, we find
\begin{eqnarray}\label{brstbreking6}
  \p_\mu\widetilde{A}_\mu &=& \p_\mu D_\mu(C)\omega~=~0\;.
\end{eqnarray}
 showing that $\widetilde{A}_\mu$ is transverse. The
field $\widetilde{A}_\mu$ also lies very close to the boundary
$\p\Omega$. However, as it follows from Gribov's original
statement\footnote{ For the benefit of the reader we
quote here Gribov's statement, proven in \cite{Gribov:1977wm}: for each field $A_{\mu}$ belonging to the Gribov region
$\Omega$ and located near the boundary $\partial \Omega$, i.e.
$A_{\mu}= C_{\mu}+a_{\mu}$, there exists an equivalent field
$\widetilde{A}_\mu$, $\widetilde{A}_\mu=C_{\mu}+a_{\mu}
+D_{\mu}(C)\omega$, near the boundary $\partial \Omega$, located,
however, on the other side of the horizon, outside of the region
$\Omega$.} \cite{Gribov:1977wm}, it is located on the side of the
horizon opposite to that of the field $A_{\mu}$, i.e. it lies
outside of the Gribov region $\Omega$.
\end{itemize}
 We can conclude thus that any infinitesimal
transformation of a gauge field configuration which belongs to the
Gribov region $\Omega$, results in another configuration which
lies outside $\Omega$. Since the BRST transformation of the gluon
field is naturally obtained from the infinitesimal gauge
transformations, it is apparent that the breaking of the BRST
symmetry looks almost as a natural reflection of   the
previous result.\\

We can also offer a pictorial depiction of what is happening. We
recall that the Gribov region $\Omega$ is convex,
bounded in all directions in field space, that every gauge field
has an equivalent representant within $\Omega$, that the origin
$A_\mu=0$ belongs\footnote{This means that perturbation theory
 belongs to $\Omega$.} to $\Omega$
\cite{Zwanziger:1982na,Dell'Antonio:1989jn,Dell'Antonio:1991xt},
 and that every gauge configuration near the horizon
$\p\Omega$ has a copy on the other side of $\p\Omega$
\cite{Gribov:1977wm}. The first 4 quoted properties are important
to make $\Omega$ a suitable domain of  integration in
the path integral, i.e. we can restrict the whole space of
$A_\mu$-configurations to $\Omega$ as proposed by Gribov. However,
  implementing this restriction in $A_\mu$-space
 jeopardizes the BRST invariance. As we have seen, if
we move throughout $A_\mu$ space with a BRST transformation (cfr
infinitesimal gauge transformations), we must unavoidably cross
the horizon $\p\Omega$. Hence, restricting the fields within the
horizon breaks the BRST invariance.

\subsection{The Maggiore-Schaden construction revisited}
The authors of the paper \cite{Maggiore:1993wq} attempted to
interpret the BRST breaking as a kind of \textit{spontaneous}
symmetry breaking. We shall now re-examine this proposal and
conclude that, instead, the BRST breaking has to be considered as an
\textit{explicit} symmetry breaking, where we shall present a few
arguments which have not been considered in \cite{Maggiore:1993wq}.
Although this discussion might seem to be only of a rather academic
interest, there is nevertheless a big difference between a
\textit{spontaneously} or \textit{explicitly} broken continuous
symmetry, since only in the former case a Goldstone mode would
emerge. For the benefit of the reader, we shall
 first explain in detail the approach of \cite{Maggiore:1993wq}.  One starts by adding the following BRST
exact term to the Yang-Mills action:
\begin{eqnarray}
    S_1 &=& s \int \d^4 x \left( \overline{c}^{a} \p_{\mu}A^a_{\mu} +
    \overline{\omega}_{\mu}^{ac} \p_{\nu} D_{\nu}^{ab} \varphi^{bc}_{\mu} \right)
    \;, \label{mg1}
\end{eqnarray}
with $s$, the same nilpotent BRST  operator as defined in
\eqref{BRST}. The first term represents the   Landau
gauge fixing, while the second term is a BRST exact piece in the
fields $(\varphi,\omega, \overline{\varphi},\overline{\omega})$.
 Of course, from expression \eqref{mg1}, it follows
that $s$ defines a symmetry of the action $S_{YM} + S_1 $.
 As a consequence,  the nilpotent operator
$s$ allows us to defines two doublets $(\varphi,\omega)$ and
$(\overline{\varphi},\overline{\omega})$. This  doublet
structure implies that we can exclude these fields from the physical
subspace \cite{Piguet:1995er,Henneaux:1992ig}, which makes $S_{YM} +
S_1 $ equivalent to the ordinary Yang Mills gauge theory. Next,
Maggiore and Schaden introduced a set of shifted fields, which
-translated to our conventions- are given by:
\begin{eqnarray}
\varphi^{ab}_{\mu} &=& \varphi^{\prime ab}_{\mu} + \gamma^2 \delta^{ab} x_{\mu} \;, \nonumber\\
\overline{\varphi}^{ab}_{\mu} &=& \overline{ \varphi}^{\prime ab}_{\mu} + \gamma^2 \delta^{ab} x_{\mu}\;, \nonumber\\
\overline{c}^{ a} &=& \overline{c}^{\prime a} + g \gamma^2 f^{abc} \overline{\omega}^{bc}_{\mu} x_{\mu} \;,\nonumber\\
b^{a} &=& b^{ \prime a} + g \gamma^2 f^{abc} \overline{\varphi}^{bc}_{\mu} x_{\mu}\;.
\end{eqnarray}
 All fields $(\varphi^{\prime ab}_{\mu}, \overline{
\varphi}^{\prime ab}_{\mu}, \overline{c}^{\prime a}, b^{ \prime
a}) $ have vanishing vacuum expectation value (VEV), namely
\begin{equation}
\langle \varphi^{\prime ab}_{\mu} \rangle = \langle \overline{
\varphi}^{\prime ab}_{\mu} \rangle = \langle \overline{c}^{\prime
a} \rangle = \langle b^{ \prime a} \rangle = 0 \;. \label{mg2}
\end{equation}
Along with these new fields ($\varphi^{\prime ab}_{\mu}$,
$\overline{\varphi}^{\prime ab}_{\mu}$, $\overline{c}^{\prime a}$,
$b^{\prime a}$),  one introduces a modified nilpotent
BRST operator $\widetilde{s}$ given by:
\begin{align}
\widetilde{s} \;\overline{c}^{\prime a} &=b^{\prime a}\;,&   \widetilde{s} b^{\prime a}&=0\;,  \nonumber \\
\widetilde{s}\varphi _{\mu}^{\prime ab} &=\omega _{\mu}^{ab}\;, & \widetilde{s} \overline{\varphi }^{\prime ab}_\mu &=0\;,  \nonumber \\
\widetilde{s} A_{\mu}^a &= -D_{\mu}^{ab} c^b \;,&  \widetilde{s} \omega_{\mu}^{ab} &= 0\;,
\end{align}
which looks exactly like \eqref{BRST}. However, we emphasize that
by introducing these new fields, the BRST operator $\widetilde{s}$
will give rise to an explicit $x$-dependence when acting on the
field $\overline{\omega}_{\mu}^{ab}$:
\begin{eqnarray}
\widetilde{s} \overline{\omega}_{\mu}^{ab} &=&
\overline{\varphi}^{\prime ab}_{\mu} + \gamma^2 \delta^{ab} x_\mu
\;. \label{mg3}
\end{eqnarray}
Furthermore, by taking the vacuum expectation value of
both sides of equation \eqref{mg3}, one gets
\begin{equation}
\langle \widetilde{s}\; \overline{\omega}_{\mu}^{ab} \rangle =
\gamma^2 \delta^{ab} x_\mu \;, \label{mg4}
\end{equation}
from which the authors of \cite{Maggiore:1993wq} infer that the BRST
operator $\widetilde{s}$ suffers from spontaneous symmetry
breaking.  Notice also that \eqref{mg4} gives a VEV to a quantity with a free Lorentz index.\\

With the introduction of the shifted
fields, we can rewrite the action $S_1$ as:
\begin{eqnarray}
    S_1 &=& \widetilde{s} \int \d^4 x \left( \overline{c}^{\prime a} \p_{\mu}A^a_{\mu} + \overline{\omega}_{\mu}^{ac} \p_{\nu} D_{\nu}^{ab} \varphi^{\prime bc}_{\mu} + g \gamma^2 f^{abc} \overline{\omega}^{bc}_{\nu} x_{\nu} \p_{\mu} A^a_{\mu} + \gamma^2 \overline{\omega}^{ac}_{\mu} \p_{\nu} D_{\nu}^{ab} \delta^{bc} x_{\mu}  \right) \;.
\end{eqnarray}
The last two terms can be simplified, leading to
\begin{eqnarray}\label{actionbrst}
    S_1 &=& \widetilde{s} \int \d^4 x \left( \overline{c}^{\prime a} \p_{\mu}A^a_{\mu} + \overline{\omega}_{\mu}^{ac} \p_{\nu} D_{\nu}^{ab} \varphi^{\prime bc}_{\mu}
   - g \gamma^2 \overline{\omega}^{ab}_{\mu} f_{abc} A^c_{\mu}  \right) \;.
\end{eqnarray}
If we calculate this action explicitly, we recover the original
Gribov-Zwanziger action, without the constant part $4\gamma^4 (N^2
-1)$. For this reason  one adds $ - \gamma^2 \widetilde{s} \int
\d^4 x \p_{\nu} \overline{\omega}^{aa}_{\mu}$ to the
action $S_1$. Doing so,  one finds
\begin{eqnarray}
    S_1 &=& \widetilde{s} \int \d^4 x \left( \overline{c}^{\prime a} \p_{\mu}A^a_{\mu} + \overline{\omega}_{\mu}^{ac} \p_{\nu} D_{\nu}^{ab} \varphi^{\prime bc}_{\mu}
   - g \gamma^2 \overline{\omega}^{ab}_{\mu} f_{abc} A^c_{\mu}  - \gamma^2 \p_{\mu} \overline{\omega}^{aa}_{\mu} \right) \nonumber\\
   &=& \int \d^4 x \left[ b^{\prime a} \p_{\mu} A^a_{\mu} + \overline{c}^{\prime a} \p_{\mu} \left(D_{\mu}^{ab} c^b \right) \right] + \int \d^4 x \left[ \overline{\varphi}^{\prime ac}_{\mu} \p_\nu D_{\nu}^{ab} \varphi^{ \prime bc}_{\mu} + \gamma^2  x_{\mu}   \p_\nu D_{\nu}^{ab} \varphi^{\prime ba}_{\mu} +  \overline{\omega}^{ac}_{\mu} \p_{\nu} \left(g  f^{akb} D^{kd}_{\nu} c^d \varphi^{\prime bc}_{\mu}  \right) - \overline{\omega}^{ac}_{\mu} \p_{\nu} D^{ab}_{\nu} \omega^{bc}_{\mu} \right] \nonumber\\
   && + \int \d^4 x\left[ -g \gamma^2 \overline{\varphi}^{\prime ab}_{\mu} f_{abc} A^c_{\mu} - g \gamma^2 \overline{\omega}^{ab}_{\mu} f_{abc} D^{cd}_{\mu} c^d - 4 \gamma^4 (N^2 - 1)\right]\;.
\end{eqnarray}
If we naively assume that we can perform a partial integration, we find after dropping the surface terms,
\begin{eqnarray}\label{sec}
S_{YM}+ S_1&=& \eqref{s1eq1} -  \; g \gamma^2 f^{abc}
\int \d^4x \; \overline{\omega}_{\mu}^{ab} D_{\mu}^{cd} c^d \;.
\end{eqnarray}
The last expression reveals that one has recovered the
Gribov-Zwanziger action from an exact $\widetilde{s}$-variation with
the addition of an extra term $\left( \;- g \gamma^2 f^{abc} \int
\d^4x \; \overline{\omega}_{\mu}^{ab} D_{\mu}^{cd} c^d \right)$.
However, this term is irrelevant as we shall explain now. Assume
that we want to compose an arbitrary Feynman diagram without any
external $\overline{\omega}$ leg and thereby using the action
\eqref{sec}. The second term from this action can never contribute
to this Feynman diagram as it contains an external
$\overline{\omega}$. Indeed, this leg requires an $\omega$-leg,
which in its turn is always accompanied by an $\overline{\omega}$
leg. Hence, the action \eqref{sec} is equivalent to the standard
Gribov-Zwanziger action \eqref{s1eq1} when we exclude the diagrams
containing external $\overline{\omega}$ legs\footnote{For our
purposes these diagrams are
irrelevant, e.g.~the vacuum energy, the gluon and ghost propagator,\ldots .}. \\

Although at first sight this construction might seem useful, it
turns out that a few points have been overlooked. Let us investigate
this in more detail. Firstly, we  point out that rather delicate
assumptions have been made concerning the partial integration. To
reveal the obstacle, we perform once more the partial integration
explicitly,
\begin{eqnarray}\label{surface}
\int \d^4 x \gamma^2 x_{\mu} \p_{\nu} D_{\nu}^{ab}
\varphi^{\prime ba}_{\mu} &=& \text{surface term} - \int \d^4 x \gamma^2
\delta_{\mu\nu} D^{ab}_{\nu} \varphi^{\prime ba}_{\mu} \;.
\end{eqnarray}
Normally, one drops the surface terms, as the fields vanish at
infinity. However in this case, as $x_{\mu}$ does not vanish at
infinity, it is not sure if the surface terms $ \propto
x_\mu$ will be zero. One would have to impose extra conditions on
the fields to justify the dropping of the surface terms. On the
other hand,  when we do not perform the partial
integration to avoid the surface terms, we are facing an explicit,
unwanted $x$-dependence in the action, resulting in an explicit
breaking of translation invariance. \\
Another way of looking at the problem consists of performing a
partial integration on the second term of the action
\eqref{actionbrst} \textit{before} applying the BRST variation
$\widetilde{s}$. Doing so, we find,
\begin{eqnarray}\label{actionbrst2}
    S_1 &=& \widetilde{s} \int \d^4 x \left( \overline{c}^{\prime a} \p_{\mu}A^a_{\mu} -  \p_{\nu} \overline{\omega}_{\mu}^{ac}  D_{\nu}^{ab} \varphi^{\prime bc}_{\mu}
   - g \gamma^2 \overline{\omega}^{ab}_{\mu} f_{abc} A^c_{\mu} - \gamma^2 \p_{\mu} \overline{\omega}^{aa}_{\mu} \right) \;.
\end{eqnarray}
Subsequently, applying the BRST variation gives,
\begin{eqnarray}
    S_1 &=& \int \d^4 x \left[ b^{\prime a} \p_{\mu}A^a_{\mu} + \overline{c}^{\prime a} \p_{\mu} \left(D_{\mu}^{ab} c^b \right) \right] + \int \d^4 x \left[- \p_\nu \overline{\varphi}^{\prime ac}_{\mu}  D_{\nu}^{ab} \varphi^{\prime bc}_{\mu} - \gamma^2 \delta^{\mu\nu}   D_{\nu}^{ab} \varphi^{\prime ba}_{\mu} - \left( \p_{\nu} \overline{\omega}^{ac}_{\mu} \right)  g  f^{akb} D^{kd}_{\nu} c^d \varphi^{\prime bc}_{\mu}  + \left( \p_{\nu} \overline{\omega}^{ac}_{\mu} \right)  D^{ab}_{\nu} \omega^{bc}_{\mu} \right] \nonumber\\
   && + \int \d^4 x\left[ -g \gamma^2 \overline{\varphi}^{\prime ab}_{\mu} f_{abc} A^c_{\mu} - g \gamma^2 \overline{\omega}^{ab}_{\mu} f_{abc} D^{cd}_{\mu} c^d - 4 \gamma^4 (N^2 - 1)\right] \nonumber\\
   &=& \eqref{s1eq1} - g \gamma^2 f^{abc} \int \d^4x\overline{\omega}_{\mu}^{ab} D_{\mu}^{cd} c^d \;.
\end{eqnarray}
In this case, we do not encounter the problem of nonvanishing
surface terms. To recapitulate, if we first let the BRST variation
act on the action \eqref{actionbrst}, and then perform a partial
integration, we find a different result than performing these two
operations the other way around. This difference is exactly given by
the surface term from equation \eqref{surface}. This discrepancy
arises of course from the explicit $x$-dependence introduced in the
BRST transformation $\widetilde{s}$, giving nontrivial
contributions. For example, we introduced a term $ - \gamma^2
\widetilde{s} \int \d^4 x \p_{\nu} \overline{\omega}^{aa}_{\mu}$
which might seem to be zero since we are looking at the integral of
a complete derivative (thus usually taken to be a vanishing surface
term), but when the BRST variation is taken first, a nontrivial
integrated piece remains.\\
Apparently, to find the correct Gribov-Zwanziger action with the
Maggiore-Schaden argument, there is some kind of a ``hidden working
hypothesis'' that \eqref{actionbrst2} is the correct action to start
with, and that partial integration is not always allowed\footnote{If
it would be allowed, one would be able to cross from the second
action \eqref{actionbrst2} to the first one \eqref{actionbrst}, but as we have
just shown, these two starting actions are inequivalent. }. The fact
that there seems to be a kind of ``preferred'' action to start with,
is just a signal that there is a problem with the boundary
conditions for some of the fields and hence surface terms when
integrating.\\

Even if  one forgets about the previous criticism, a second problem
arises. In the Gribov-Zwanziger approach, we recall that the
parameter $\gamma$ is not free and is determined by the horizon
condition \eqref{gapgamma}. As it  has been explained in section
\ref{GZ}, the solution $\gamma =0$ is excluded. In equation
\eqref{perturbative}, we found a solution for $\gamma \not= 0$. This
gave rise to a positive vacuum energy $E_{\mathrm{vac}} > 0$, as one
can see from equation \eqref{positiveenergy}, see also
\cite{Dudal:2005}. However, according to Maggiore-Schaden argument,
at one loop  order the stable solution should be that corresponding
to $\gamma=0$ \cite{Maggiore:1993wq}, as, if $\gamma =0$, the vacuum
energy would be vanishing, i.e. $E_{\mathrm{vac}} = 0$, which is
energetically favored over a positive vacuum energy. This delivers a
contradiction with the Gribov-Zwanziger approach,  as the
restriction to the Gribov region
requires that $\gamma \not= 0$, thus giving a positive energy $E_{\mathrm{vac}} > 0$ at one loop.\\

To end this section, let us now consider the new operator $\int \d^4
x \left( \overline{\varphi}^{ab}_{\mu} \varphi^{ab}_{\mu} -
\overline{\omega}^{ab}_{\mu} \omega^{ab}_{\mu} \right)$ within the
Maggiore-Schaden approach. We observe that we obtain an explicit
$x$-dependence if we rewrite this operator in terms of the new
fields,
\begin{eqnarray}\label{massaop}
\int \d^4 x \left( \overline{\varphi}^{ab}_{\mu} \varphi^{ab}_{\mu}
- \overline{\omega}^{ab}_{\mu} \omega^{ab}_{\mu} \right) &=& \int \d^4 x
\left( \overline{\varphi}^{\prime ab}_{\mu} \varphi^{\prime ab}_{\mu}
- \overline{\omega}^{ab}_{\mu} \omega^{ab}_{\mu}
+ \gamma^2 x_{\mu} \varphi^{\prime aa}_{\mu}
- \gamma^2 x_{\mu} \overline{\varphi}^{\prime aa}_{\mu}
- \gamma^4 x_{\mu} x_{\mu} (N^2-1) \right) \;.
\end{eqnarray}
However, this $x$-dependence is necessary so that \eqref{massaop}
would be invariant under the new BRST symmetry $\widetilde{s}$,
\begin{eqnarray}
\widetilde{s} \int \d^4 x \left( \overline{\varphi}^{\prime ab}_{\mu} \varphi^{\prime ab}_{\mu} - \overline{\omega}^{ab}_{\mu} \omega^{ab}_{\mu} + \gamma^2 x_{\mu} \varphi^{\prime aa}_{\mu} - \gamma^2 x_{\mu} \overline{\varphi}^{\prime aa}_{\mu} - \gamma^4 x_{\mu} x_{\mu} (N^2-1) \right) &=& 0 \;.
\end{eqnarray}
A second option is to introduce the BRST $\widetilde s$-exact mass
operator
\begin{eqnarray}
\widetilde s \int \d^4 x \left( \overline{\omega}_{\mu}^{ab} \varphi_{\mu}^{ab} \right) &=& \int \d^4 x \left(  \overline{\varphi}^{\prime ab}_{\mu} \varphi^{ ab}_{\mu} - \overline{\varphi}^{ab}_{\mu} \varphi^{ ab}_{\mu}  + \gamma^2 x_{\mu} \varphi_{\mu}^{aa}  \right) \;,
\end{eqnarray}
which also displays an explicit $x$-dependence.\\
Finally, let us consider a third and last possible option. If we would have started with the following mass operator,
\begin{eqnarray}
\int \d^4 x \left( \overline{\varphi}^{\prime ab}_{\mu} \varphi^{\prime ab}_{\mu} - \overline{\omega}^{ab}_{\mu} \omega^{ab}_{\mu} \right) \;,
\end{eqnarray}
which does not contain an $x$-dependence, this operator is not left invariant by the symmetry $\widetilde{s}$. In fact,
\begin{eqnarray}
\widetilde{s} \int \d^4 x \left( \overline{\varphi}^{\prime ab}_{\mu} \varphi^{\prime ab}_{\mu}
- \overline{\omega}^{ab}_{\mu} \omega^{ab}_{\mu} \right) &=& - \gamma^2
\int \d^4 x\ x_{\mu} \omega^{aa}_{\mu} \;.
\end{eqnarray}

One sees that with the introduction of the new mass
operator, the Maggiore-Schaden construction will always give rise
to an explicit breaking of translation invariance if the BRST invariance $\widetilde s$ has to be preserved. We thus
conclude that the Maggiore-Schaden construction cannot be
implemented in the presence of the new operator and even without
the new mass operator we have collected a few arguments
 from which the frame of  a possible
spontaneous symmetry breaking cannot be applied  to the
Gribov-Zwanziger action.

\subsection{ A few remarks on the Kugo-Ojima confinement criterion}
In this section we shall take a closer look at the Kugo-Ojima
confinement criterion \cite{Kugo:1979gm} in relation to the
Gribov-Zwanziger action. In the literature, it is
usually stated that the Kugo-Ojima confinement criterion is
realized when the Gribov-Zwanziger scenario is realized. A key
ingredient in the criterion is   $u(0) = -1$, whereby
 $u(0)$ is the value at zero momentum of a specific
Green function. $u$ is related to the ghost propagator in the Landau
gauge according to \cite{Kugo:1995km}
\begin{eqnarray}
  \; \mathcal{G}(p^2)_{p^2 \approx 0} &=& \frac{1}{p^2}
\frac{1}{1+u(p^2)} \;.
\end{eqnarray}
From this expression, it is obvious that an infrared enhanced ghost
propagator results in  $u(0) = -1$, thereby fulfilling the
criterion. Let us recall here that the derivation of the Kugo-Ojima
criterion is based on the assumption of an exact BRST invariance and
is written down in a Minkowskian rather than an Euclidean
space-time. This has a few repercussions:
\begin{itemize}
\item At a  nonperturbative level,  some
care should be taken when passing from Euclidean to Minkowski
space-time. According to our understanding, it is not clear whether
a Wick rotation can always be implemented. E.g., the gluon
propagator \eqref{wishedform} can exhibit two complex conjugate
poles, so one should be careful  of not crossing these poles when
the contour is Wick rotated. Clearly, there could be potential
caveats when considering a more complicated gluon propagator. \item
A more crucial shortcoming is the following. As we have emphasized
in the foregoing section, the restriction to the Gribov region
inevitably leads to a breaking of the BRST symmetry  which, however,
was the very starting point of the Kugo-Ojima analysis. In
addition, parts of the Kugo-Ojima study rely on analyzing the charge
of the global color current and the expression of a piece of it in
terms of the BRST symmetry generator. In our opinion, as the
Gribov-Zwanziger action is essentially different from the usual
Faddeev-Popov fixed (Landau gauge) action due to new fields, extra
interactions and especially another symmetry content, the Kugo-Ojima
analysis cannot simply be applied to the Gribov-Zwanziger formalism,
although both might superficially seem to be in accordance with each
other. Therefore, it seems to us that one cannot verify the
Kugo-Ojima criterion ($u(0)=-1$) when the restriction to the Gribov
horizon is taken into account\footnote{This would also include
Schwinger-Dyson results which implemented the restriction to the
Gribov region by suitable boundary conditions.}.
\item The latest lattice data point towards a ghost propagator which is no longer
enhanced, so that the condition  $u(0)=-1$ does not seem to be realized anyhow.
\end{itemize}

\section{Discussion}
Our starting point was the original localized Gribov-Zwanziger action, $S_{GZ}$, and the
observation of the new lattice data, which shows an infrared suppressed, positivity
violating gluon propagator, nonvanishing at the origin and a ghost propagator
which is no longer enhanced. However, the propagators  corresponding to the
original Gribov-Zwanziger action are not in accordance with these new lattice data. Hence, we
have searched for a solution by looking at nonperturbative effects like condensates. Therefore, we have
added two extra terms to the Gribov-Zwanziger action, $S_M = M^2 \int  \d^4x\left[\left(\overline{\varphi}\varphi-\overline{\omega}\omega\right) + \frac{2 (N^2 -1)}{ g^2 N}   \varsigma  \lambda^2  \right]$ . A first intuitive argument why we added the first term, $M^2 \int  \d^4x\left(\overline{\varphi}\varphi-\overline{\omega}\omega\right)$, was the following. In the Gribov-Zwanziger action, $S_{GZ}$, an $A\varphi$-coupling is already present at the quadratic level. Therefore, altering the $\varphi$-sector will be translated to the $A$-sector, thus modifying the gluon propagator. Secondly, this condensate is already present perturbatively as, at lowest order, we have found
\begin{eqnarray}
\Braket{ \overline{\varphi}\varphi-\overline{\omega}\omega}  &=&
\frac{3(N^2 - 1)}{64 \pi} \lambda^2 \;,
\end{eqnarray}
with $\lambda^4= 2g^2 N \gamma^4$. This implies that the condensate is nonvanishing for $\gamma \not= 0$ already in the original Gribov-Zwanziger action. It was therefore very natural to add this operator to the theory. The second pure vacuum term, $M^2 \int  \d^4x \frac{2(N^2 -1)}{ g^2 N} \varsigma  \lambda^2$, was added in order to stay within the horizon or equivalently, to keep $\sigma(0)$ smaller than 1 when the horizon condition is implemented. We have fixed $\varsigma$ by imposing $\left.\frac{\partial \sigma(0)}{\partial M^2} \right|_{M^2=0}= 0$; this ensures a smooth limit to the original Gribov-Zwanziger action.\\

The extended Gribov-Zwanziger action, $S_{GZ} + S_M$, has many
interesting features. Not only is this action renormalizable, it is
also remarkable that no new  renormalization factors are
necessary for the proof of its renormalizability, meaning that only
two independent  renormalization factors are required.
As an extra feature, we have also shown that $S_{GZ} + S_M + S_{A^2}$,
with $S_{A^2} = \; \frac{m^2}{2} \int \d^4x \;A_{\mu}^2$,
is renormalizable.\\

Another important observation is that the gluon propagator is
already modified at tree level. We have found
\begin{eqnarray}
\mathcal{D}(p^2) &=& \frac{p^2 + M^2}{p^4 + (M^2+m^2)p^2 +
\lambda^4+M^2m^2 } \;.
\end{eqnarray}
 This type of propagator is in qualitative agreement
with the most recent lattice data, which was the starting point of
our analysis. In section IV the gluon propagator at zero momentum
was also presented at one loop (see equation \eqref{oneloop}),
 where we switched off the effects related to $A^2$ by
setting $m^2=0$. By virtue of the novel mass $M^2$,
$\mathcal{D}(p^2) \not= 0$ at zero momentum. Also the one loop ghost
propagator is modified. At small momenta we have obtained,
\begin{eqnarray}
\mathcal{G}(p^2)_{p^2 \approx 0} &=& \frac{1}{p^2} \frac{1}{ 1- \sigma} \;.
\end{eqnarray}
with
\begin{eqnarray}
\sigma (0) &=& 1 + M^2 \frac{3 g^2 N}{64 \pi^2}  \frac{1}{ \sqrt{M^4 - 4 \lambda^4}} \left[  \ln \left(  M^2  +  \sqrt{M^4 - 4 \lambda^4} \right) - \ln \left(M^2 -  \sqrt{M^4 - 4 \lambda^4} \right) \right] - \left( \frac{3g^2 N}{128 \pi} \right) \frac{M^2}{\lambda^2}\;.
\end{eqnarray}
We see that the ghost propagator is clearly no longer enhanced, again in accordance with the most recent lattice data.\\

Up to this point,  the mass $M^2$ was put in by hand. However, we have treated $\left( \overline{\varphi}^a_i \varphi^a_{i} - \overline{\omega}^a_i \omega^a_i \right)$ as a composite operator coupled to the source $J = M^2$. In this way,  we have been able to find nonperturbative effects induced by this composite operator without altering the original Gribov-Zwanziger action, and making the mass $M^2$ dynamical. We have developed two methods to find such nonperturbative effects. The first method uses the well known principles of the effective action formalism. Unfortunately, the calculations become intractable. Therefore, we have implemented a second method, the variational principle. Intuitively, we have included effects of the mass term without altering the original Gribov-Zwanziger action by performing a
 suitable resummation. With the help of this technique, we have found in the $\MSbar$-scheme that $\sigma (0)$, the one loop correction to $\left(p^2 \mathcal{G}(p^2)\right)^{-1}_{p^2 \approx 0 }$  is given by\footnote{We set $\lms=0.233\mathrm{GeV}$, the value reported in \cite{Boucaud:2001st}.}
\begin{eqnarray}\label{disc1}
\sigma(0) &=& 0.93 \;,
\end{eqnarray}
resulting in a non-enhanced ghost propagator. Simultaneously, for the one loop gluon propagator at zero momentum, we have found
\begin{eqnarray}\label{disc2}
{\cal D}^{(1)}(0) = \frac{0.63}{ \lms^2}\sim \frac{11.65}{\mathrm{GeV}^2} \;,
\end{eqnarray}
which is nonzero. The corresponding value for the coupling constant is smaller than 1, see equation \eqref{koppel1},
which is acceptable for a perturbative expansion. We have also checked the positivity
violation of the gluon propagator with the help of the variational
technique and again, our results were in nice agreement with
lattice results: not only is the shape of the temporal correlator
$\mathcal{C}(t)$, displayed in FIG.~\ref{ct} in qualitative
agreement, also the  value of the point, $t \sim 1.5$ fm, at which
the violation of positivity starts is consistent with the results
reported in lattice investigations. Using the plots displayed in
\cite{Bogolubsky:2007ud} which were also obtained in the $SU(3)$
case, one can extract a rough lattice estimate for the quantities
\eqref{disc2} and \eqref{disc1},
\begin{eqnarray}\label{disc2bis}
 {\cal D}^{\mathrm{lattice}}(0) &\sim& \frac{13}{\mathrm{GeV}^2} \;.
\end{eqnarray}
\begin{eqnarray}\label{disc1bis}
 p^2 \mathcal{G}(p^2)^{\mathrm{lattice}}_{p^2 \approx 0} ~\sim~ 5  ~\Leftrightarrow~ \sigma(0)^{\mathrm{lattice}} &\sim& 0.8 \;,
\end{eqnarray}
We notice that our lowest order approximations \eqref{disc2} and \eqref{disc1} are qualitatively compatible with the current lattice values.\\

To conclude, we would like to emphasize that the original
Gribov-Zwanziger action already breaks the BRST symmetry. Due to
this breaking, it is in unclear at present how to define the
observables of the theory in the nonperturbative infrared region.
According to our understanding, this breaking cannot be interpreted
as a spontaneous breaking, according to the proposal of
\cite{Maggiore:1993wq}. In fact, we have argued that the BRST
breaking is a natural consequence of introducing the restriction to
the Gribov region. In addition, we have underlined that the
presence of the BRST breaking term in the Gribov-Zwanziger action
provides a consistent way to ensure that the restriction to the
Gribov region can have physical consequences, i.e. that the Gribov
parameter $\gamma$  enters the expectation value of physical, gauge
invariant correlators. In the absence of such a breaking term, the
Gribov mass parameter would play the role of an unphysical gauge
parameter. The presence of the breaking is thus a necessary tool
within the Gribov-Zwanziger approach, allowing for the introduction
of a nonperturbative mass parameter in a local and renormalizable
way. Finally, we have also commented on the Kugo-Ojima confinement
criterion. Since it is fundamentally based on the concept of an
exact BRST symmetry, it cannot be straightforwardly related to the
Gribov-Zwanziger framework
due to the breaking.\\

In summary, this paper presented the 4D analysis of the gluon and
the ghost propagator within the Gribov-Zwanziger framework. By
comparing these results with recent lattice data, we have found a
good qualitative agreement. The ghost and gluon propagator have
also been extensively studied on the lattice in 2 and 3 dimensions
\cite{Cucchieri:2007md,Cucchieri:2007rg,Cucchieri:2008fc,Maas:2007uv}.
The 3D and 2D analysis of the extended Gribov-Zwanziger action,
and a comparison with the lattice data, is  currently under
consideration.

\section*{Acknowledgments}
We thank D. Zwanziger and M. Schaden for useful discussions. The
Conselho Nacional de Desenvolvimento Cient\'{i}fico e
Tecnol\'{o}gico (CNPq-Brazil), the Faperj, Funda{\c{c}}{\~{a}}o de
Amparo {\`{a}} Pesquisa do Estado do Rio de Janeiro, the SR2-UERJ
and the Coordena{\c{c}}{\~{a}}o de Aperfei{\c{c}}oamento de Pessoal
de N{\'{i}}vel Superior (CAPES) are gratefully acknowledged for
financial support.  D.~Dudal and N.~Vandersickel acknowledge the
financial support from the Research Foundation - Flanders (FWO).

\appendix*
\section{The one loop effective potential}

We explain in  detail how we obtained equation
\eqref{effectiveenergy}. We start by evaluating the integral
appearing in $W(J)$. We recall that this integral originates from:
\begin{eqnarray}
 &&  \; \int \d A_{\mu} \exp - \frac{1}{2} \int \d^4x
A_{\mu}^a (\Delta^{ab}_{\mu\nu}) A^b_{\nu} = \left[\det \left[ -
\left( \p^2 + \frac{2g^2  N \gamma^4}{\p^2 - M^2}\right)
\delta_{\mu\nu} - \p_{\mu} \p_{\nu} \left( 1 -
\frac{1}{\alpha}\right)  \right] \right]^{-1/2} = \e^{-\frac{1}{2}
\Tr \ln Q_{\mu\nu}^{ab}}\;,
\end{eqnarray}
with $ Q_{\mu\nu}^{ab} = - \left( \p^2 +   \frac{2g^2  N
\gamma^4}{\p^2 - M^2}\right) \delta_{\mu\nu} - \p_{\mu} \p_{\nu}
\left( 1 - \frac{1}{\alpha}\right) $. From this expression
 it follows that we need to calculate $\frac{1}{2} \Tr
\ln Q_{\mu\nu}^{ab}$ to obtain the energy functional:
\begin{eqnarray}
\frac{1}{2} \Tr \ln Q_{\mu\nu}^{ab} &=& \frac{N^2 - 1}{2} (d-1) \Tr \ln \left(- \p^2 - \frac{2g^2 N \gamma^4}{ \p^2 - M^2}\right) \nonumber\\
&=&  \frac{N^2 - 1}{2} (d-1) \left\{ \Tr \ln \left(- \p^2(\p^2 -M^2)
- 2g^2 N \gamma^4 \right\}   - \Tr \ln (-\p^2 + M^2) \right\}\,.
\end{eqnarray}
The second part is a standard integral and evaluated as:
\begin{eqnarray}
\Tr \ln (-\p^2 + M^2) &=& \frac{-\Gamma(-d/2)}{(4\pi)^{d/2}} \frac{1}{(M^2)^{-d/2}}\;,
\end{eqnarray}
with $\Gamma$ the Euler Gamma-function. Using dimensional
regularization, $d= 4- \epsilon$ we obtain,
\begin{eqnarray}\label{f1}
 -\frac{N^2 - 1}{2} (d-1) \Tr \ln (-\p^2 + M^2) &=& -3 \frac{N^2 - 1}{64 \pi^2} M^4 \left( -\frac{5}{6} - \frac{2}{\epsilon} + \ln \frac{M^2}{\overline{\mu}^2}
 \right)\;.
\end{eqnarray}
We recall that we work in the $\MSbar$ scheme. Next, we try to convert the first part in to the standard form,
\begin{align}\label{f2}
\frac{N^2 - 1}{2} (d-1)  \Tr \ln \bigl(-& \p^2(\p^2 -M^2) - 2g^2 N \gamma^4 \bigr) \nonumber\\
=& \frac{N^2 - 1}{2} (d-1) \Tr \ln \left( -\p^2 + m_1^2 \right) + \Tr \ln \left( -\p^2 + m_2^2 \right) \nonumber\\
=& \frac{N^2 - 1}{2} (d-1) \left[\frac{-\Gamma(-d/2)}{(4\pi)^{d/2}} \frac{1}{(m_1^2)^{-d/2}} + \frac{-\Gamma(-d/2)}{(4\pi)^{d/2}} \frac{1}{(m_2^2)^{-d/2}} \right]\nonumber\\
=&\frac{N^2 - 1}{2} (d-1)\left[ \frac{-\Gamma(-d/2)}{(4\pi)^{d/2}} \frac{1}{(m_1^2)^{-d/2}} + \frac{-\Gamma(-d/2)}{(4\pi)^{d/2}} \frac{1}{(m_2^2)^{-d/2}} \right]\nonumber\\
=&  3 \frac{N^2 - 1}{64 \pi^2} \left( m_1^4 \left( -\frac{5}{6} -
\frac{2}{\epsilon} + \ln \frac{m_1^2}{\overline{\mu}^2} \right)  +
m_2^4 \left( -\frac{5}{6} - \frac{2}{\epsilon} + \ln
\frac{m_2^2}{\overline{\mu}^2} \right) \right)+O(\epsilon) \;,
\end{align}
where we have used the notational shorthand \eqref{notationalshorthand}. We still have to calculate the first and the second term of \eqref{effectiveenntuitgerekend}. For the first term, we recall that
\begin{eqnarray}
\gamma_0^4 &=& Z_{\gamma^2}^2 \gamma^4\;,  \qquad \text{with} \qquad
Z_{\gamma^2}^2 = 1+ \frac{3}{2} \frac{g^2N}{16 \pi^2}
\frac{1}{\epsilon}\;,
\end{eqnarray}
with $Z_{\gamma^2}$ defined in \eqref{zm}, so we find
\begin{eqnarray}\label{f3}
-d(N^2 - 1)\gamma^4_0 &=& -4 (N^2 - 1)\gamma^4 - 4 \frac{3}{2} (N^2 - 1)\frac{g^2 N}{16\pi^2}\frac{1}{\epsilon} \gamma^4 + \frac{3}{2} \frac{g^2N}{16 \pi^2}\gamma^4 (N^2-1) \;.
\end{eqnarray}
The second term is invariant under renormalization and therefore given by
\begin{eqnarray}
 \frac{d (N^2 -1)}{ g^2 N}  \varsigma \ \lambda^2 J.
\end{eqnarray}
From equation \eqref{f1}, \eqref{f2} and \eqref{f3} we see that the
infinities cancel out nicely, so that the functional energy reads,
\begin{eqnarray}
W^{(1)}(J) &=& - \frac{4 (N^2 - 1)}{2 g^2 N} \lambda^4 + \frac{d (N^2 -1)}{ g^2 N}  \varsigma \ \lambda^2 J + \frac{3(N^2 - 1)}{64 \pi^2} \left( \frac{8}{3} \lambda^4 + m_1^4 \ln \frac{m_1^2}{\overline{\mu}^2}
+ m_2^4 \ln \frac{m_2^2}{\overline{\mu}^2} - J^2 \ln \frac{J}{\overline{\mu}^2}\right) \;,
\end{eqnarray}
which is exactly expression \eqref{effectiveenergy}.

\end{document}